\documentclass[11pt]{article}
\usepackage{epsfig,amscd,amssymb}
\begin{document}

%
%
%
%
\title{Integrable sigma models and perturbed coset models}

\author{Paul Fendley\\
Department of Physics\\
University of Virginia\\
Charlottesville, VA 22904-4714\\
{\tt fendley@virginia.edu}
}
\maketitle

\begin{abstract}

Sigma models arise frequently in particle physics and condensed-matter
physics as low-energy effective theories. In this paper I compute the
exact free energy at any temperature in two hierarchies of integrable
sigma models in two dimensions. These theories, the $SU(N)/SO(N)$
models and the $O(2P)/O(P)\times O(P)$ models, are asymptotically free
and exhibit charge fractionalization.  When the instanton coupling
$\theta=\pi$, they flow to the $SU(N)_1$ and $O(2P)_1$ conformal field
theories, respectively.  I also generalize the free energy computation
to massive and massless perturbations of the coset conformal field
theories $SU(N)_k/SO(N)_{2k}$ and $O(2P)_k/O(P)_k \times O(P)_k$.

\end{abstract}

\section{Introduction}

Two-dimensional sigma models
have been the subject of a huge amount of study because they are
interesting toy models for gauge theories, because they often arise in
experimentally-realizable condensed-matter systems, because this is
the highest dimension in which they are naively renormalizable, and
because of the powerful theoretical methods applicable.

One of the nice things about sigma models is that the same model
can often describe completely different physics. The reason is
that in many situations, the precise sigma model of interest
follows  mainly (or sometimes
entirely) from the symmetries. For example, sigma models
often arise in theories
of interacting fermions invariant under some group $G$. 
If some fermion bilinear gets an expectation value manifestly
invariant under some subgroup $H$, then the excitations at low energy
can be described by a field taking values in $G/H$. Put another
way, the expectation value gives the fermions mass at some scale $M$.
One can then integrate out fermionic excitations, leaving
only bosonic $G/H$ excitations with masses below $M$. The sigma
model describes the interactions of these low-energy excitations,
and is independent of many of the details of the
original theory. This is why vastly different theories may end up 
having the same low-energy physics.

Two-dimensional $G/H$ sigma models all have a global symmetry group
$G$, even though the fields take values in the smaller space $G/H$.
This is one big difference between two and higher dimensions.  In
higher dimensions, the symmetry $G$ of these sigma models would be
spontaneously broken to $H$, and in the effective low-energy-theory,
the $G$ symmetry is not manifest.  In other words, in higher
dimensions the sigma model describes the physics of the massless
Goldstone bosons.  However, the Mermin-Wagner-Coleman theorem says
that in two dimensions continuous symmetries cannot be spontaneously
broken. The way these sigma models satisfy this theorem is to give the
would-be Goldstone bosons a mass and keep the original global symmetry
intact.

In particular, many interesting sigma models in two dimensions are
asymptotically free. At large energies the interactions are weak, but
at low energies the interactions are strong.  Naively, there seems to
be no mass scale in the theory (the coupling constant $g$ is
dimensionless), but a scale $\mu$ appears in the theory as a result of
short-distance effects which need to be renormalized. The coupling $g$
depends on this scale.  At $\mu$ large, $g(\mu)$ is small, so the
theory is effectively free, while as $\mu$ decreases, $g(\mu)$
increases.  In renormalization-group language, there is an unstable
trivial fixed point at $g=0$.  For $G/H$ sigma models, the manifold
$G/H$ has dimension dim$G - $dim$H$, so as $g\to 0$ the theory reduces
to dim$G - $dim$H$ free bosons.

Very elaborate techniques of perturbation theory have been developed
to describe sigma models in the regime where $g(\mu)$ is small 
(see \cite{Friedan}). However, when a sigma model is being
used as an effective theory, it is only applicable to the relevant
physics at low energies, where $\mu \ll M$. Usually in this regime,
$g(\mu)$ is large. Thus while the perturbative techniques give
valuable information, they may not tell the whole story.  To
understand the regime where $g(\mu)$ is large, one must utilize
alternative techniques. Large-$N$ expansions are a common and useful
tool. However, for most applications $N$ is small. 
For example, an application of great current interest in
the condensed matter community is in sigma models describing
disordered systems.  These sigma models are derived by using the
replica trick, which requires sending
$N\to 0$ at the end of the computation. Obviously, large-$N$
expansions are not necessarily going to be reliable here.

Luckily, for two spacetime dimensions there are other non-perturbative
methods applicable.  Many sigma models are integrable, with an
infinite number of conserved currents. The resulting conserved charges
constrain the system, making exact computations possible, even at
strong coupling.  The aim of these paper is to attempt to discuss a
number of aspects of integrable sigma models. I will derive the exact
free energy at finite temperature and in the presence of a magnetic
field. This makes it possible to compute the susceptibility and
specific heat. It also makes it possible to understand exactly the
effects of the theta term, a modification of the sigma model action
which drastically changes the low-energy physics.

One extremely interesting question is if $g(\mu)$ continues to
increase as $\mu$ decreases, or if it reaches a fixed point.  The
existence of a fixed point obviously affects the physics
enormously. In the sigma models describing disordered systems, $g$ is
related to the conductance of the system.  If there is a fixed point,
the system is a conductor, with conductance determined by the value of
$g$ at the fixed point. If there is no fixed point, the system is an
insulator.  In the former case, the excitations of the model are
massive, while in the latter, they are massless.  For the models
discussed in this paper, a non-trivial fixed point appears if a theta
term is added to the sigma model action.  The theta term has no effect
on perturbation theory.  Nevertheless, as shown in
\cite{Haldane,Pruisken,ZZtheta,metheta}, its presence can result in
the appearance of a fixed point at large $g$, completely unseen in
perturbation theory.

There are two sets of sigma models to be discussed in this paper.
Their actions can be written
conveniently in terms of a symmetric matrix field $\Phi$ as
\begin{equation}
S= \frac{1}{g}
 \hbox{tr}\, \int d^2x\ \partial^\mu \Phi^\dagger \partial_\mu \Phi
\label{action}
\end{equation}
along with the constraint
\begin{eqnarray}
\Phi^\dagger\Phi = \Phi^* \Phi= I
\label{restriction}
\end{eqnarray}
where $I$ is the identity matrix.
The constraint (\ref{restriction}) that $\Phi$ be unitary
can easily be imposed by adding a potential like
$\lambda\,\hbox{tr}\,(\Phi^\dagger \Phi - I)^2$ and taking $\lambda$ large.
In theories
with interacting fermions, this often results from introducing a
bosonic field to replace four-fermion interaction terms with Yukawa
terms (interactions between a boson and two fermions). Integrating out
the fermions then gives such a potential for the bosons
and hence the sigma model.

In the first set of models discussed in this paper,
the field takes values on the $SU(N)/SO(N)$
manifold. This corresponds to taking $\Phi$ to be a symmetric, unitary
$N\times N$ matrix of determinant $1$. 
The simplest case, $N=2$, corresponds to the manifold $SU(2)/SO(2)$ being
a two-sphere. This is because a general symmetric unitary $2\times 2$
matrix of determinant one can be written as
$$\pmatrix{v_1 + i v_2& iv_3 \cr i v_3 & v_1 - i v_2}$$ where $v_1,$
$v_2$ and $v_3$ are real and obey $(v_1)^2 + (v_2)^2 + (v_3)^2=1$.

In the second set of models discussed in this paper, the field takes
values on the $O(2P)/O(P)\times O(P)$ manifold. This corresponds to
taking $\Phi$ to be a symmetric, orthogonal, real, and traceless
$2P\times 2P$ matrix. There are several correspondences between the
two sets of models, because $SO(6)=SU(4)/{\bf Z}_2$,
$SO(4)=SU(2)\times SU(2)/{\bf Z}_2$, and $SO(3)=SU(2)/{\bf Z}_2$.  The
case $P=2$ therefore reduces to two decoupled copies of the
two-sphere, whereas the sigma model with $P=3$ is equivalent to the
$SU(4)/SO(4)$ sigma model.

The reason these $G/H$ manifolds can be described in terms of 
symmetric matrices
is as follows.
In both cases, the global symmetry $G$ acts on the field $\Phi$ as
\begin{equation}
\Phi \to U \Phi U^T
\label{symmetry}
\end{equation}
where $U$ is a unitary matrix of determinant one. This transformation
preserves the fact that $\Phi$ is a symmetric matrix with
determinant $\pm 1$. 
In the $O(2P)/O(P)\times O(P)$ sigma models, the matrix $\Phi$ is also
real. To preserve this reality, $U$ must be real as well, so $G=O(2P)$.
The eigenvalues of a orthogonal matrix must be $\pm 1$, and if
the matrix is traceless as well, there must be the same number of
$+1$ and $-1$ eigenvalues.
The field $\Phi$ in this case can
diagonalized with an orthogonal matrix $U$, so $\Phi$ can
be written 
$$\Phi=U\Lambda U^T\qquad\quad \Phi\in O(2P)/O(P)\times O(P),$$
where $U$ is in $O(2P)$, and $\Lambda$ is the matrix with
$P$ values $+1$ and $P$ values $-1$ on the
diagonal.   Different $U$ can result in
the same $\Phi$: the
subgroup leaving $\Phi$ invariant is $H=O(P)\times O(P)$. 
This is why the space of symmetric orthogonal traceless matrices
is indeed $O(2P)/O(P)\times O(P)$.
For the  $SU(N)/SO(N)$ models, $U$ can be any unitary
matrix of determinant one, so the global symmetry
$G$ is indeed $SU(N)$. Field configurations here
can be written in the form
$$\Phi=U U^T\qquad\quad\Phi\in SU(N)/SO(N)$$
where $U$ is in $SU(N)$. The subgroup $H$ leaving
$\Phi$ invariant is $SO(N)$.
For example, $\Phi=I$ for any real $U$ in $SU(N)$,
i.e.\ if $U$ is in the real subgroup $SO(N)$ of $SU(N)$.  This is
why $H=SO(N)$ here.

Under renormalization, the matrix $\Phi$ preserves its form:
e.g., it remains symmetric.
In other words, the space $G/H$
preserves its ``shape'' under renormalization, with only the overall
volume changing. The effect of renormalization is to increase
the curvature (increase $g$). These sigma models are all
asymptotically free, so going to high energies decreases $g$.
This behavior happens for all sigma models on symmetric spaces
$G/H$ (where $H$ is a maximal subgroup of $G$).

With the action (\ref{action}), there is no fixed point at large
$g$. However, if one adds a theta term, there is a non-trivial
fixed point in these sigma models \cite{metheta}.
A theta term affects field configurations 
with non-zero winding number $n$, which are called instantons.
The winding number is a topological
invariant; roughly speaking, it counts the number of times the field
configuration wraps around the two-dimensional spacetime. 
The theta term is then
\begin{equation}
S_\theta = i n\theta.
\end{equation}
If the winding number $n$ takes integer values, the theory is periodic
under shifts of the coupling $\theta$ to $\theta + 2\pi$.  This is why
the coupling $\theta$ is often called an angle. However for the general
cases considered here, $n$ can take just two values, $0$ and $1$. This
means that $\theta$ takes just two values here: $\theta = 0$ and
$\theta =\pi$. The variables $n$ and $\theta$ should be thought of as
Fourier conjugates. Adding the $\theta$ term to the action amounts to
doing a discrete Fourier transform.

For the sphere sigma model (the case $N=2$ or $P=2$ here), $n$ takes
integer values. It was argued in \cite{Haldane,Affleck} and proven in
\cite{ZZtheta} that when $\theta=\pi$ in the sphere sigma model, there
is a non-trivial fixed point at large $g$. This behavior is widely
believed to persist in other models with a $\theta$ angle (see
\cite{nato} for a review).  
An important question is therefore whether
the existence of these non-perturbative fixed points in sigma models
at $\theta=\pi$ can be generalized. In \cite{metheta}, 
it was shown that the $SU(N)/SO(N)$ and $O(2P)/O(P)\times O(P)$
sigma models have non-trivial fixed points at $\theta=\pi$.
The former fixed points are described by the $SU(N)_1$ WZW theory,
while the latter are described by the $O(2P)_1$ WZW theory.
The exact spectrum and $S$ matrices were found, and used to compute
the energy at zero temperature in the presence of a background field.
This computation essentially proves the existence of these fixed
points. 

It is the purpose of this paper to complete this proof
by studying the behavior of these models at finite temperature.
I will compute a $c$-function \cite{cthm} which clearly shows how the
field theory flows from the trivial fixed point ($g=0$)
to the non-trivial fixed point at some large value of $g$.
This computation also makes it possible to compute the
specific heat and susceptibility at both $\theta=0$
and $\theta=\pi$, a fact which will be useful
in other work \cite{robert}.

In section 2, I discuss the thermodynamic
Bethe ansatz formalism necessary to do the computation. 
In section 3, I compute the free energy at any temperature for
the massive $\theta=0$ sigma models. In section 4, I compute the
free energy for the massless $\theta=\pi$ models. In section
5, I discuss some related coset models. I conclude in section
6 by discussing the symmetries of these sigma models, and the
prospects for generalizing these results to other sigma models.

\section{The Thermodynamic Bethe Ansatz}

The proof that the sphere sigma model has a
non-trivial fixed point at $\theta=\pi$ 
utilizes the integrability of the model at $\theta=0$ and
$\pi$
\cite{ZZtheta,sausage}.
Integrability means that there are an infinite number of
conserved currents which allow one to find exactly the spectrum of
quasiparticles and their scattering matrix in the corresponding $1+1$
dimensional field theory. The quasiparticles for $\theta=0$ are gapped
and form a triplet under the $SU(2)$ symmetry \cite{ZZ},
while for $\theta=\pi$
they are gapless, and form $SU(2)$ doublets (left- and right-moving)
\cite{ZZtheta}.  This is a beautiful example of charge
fractionalization: the fields $(v_1,v_2,v_3)$ form a triplet under the
$SU(2)$ symmetry, but when $\theta=\pi$ the excitations of the system
are doublets. To prove that this is the correct particle spectrum,
first one computes a scattering matrix for these particles which is
consistent with all the symmetries of the theory. From the exact $S$
matrix, the $c$ function can be computed.  It was found that at high
energy $c$ indeed is $2$ as it should be at the trivial fixed point,
while $c=1$ as it should be at the $SU(2)_1$ low-energy fixed point
\cite{ZZtheta}.

As an even more detailed check, the free energy at
zero temperature in the presence of a magnetic field was computed for
both $\theta=0$ \cite{Hasen} and $\pi$ \cite{sausage}. The results can
be expanded in a series around the trivial fixed point. One can
identify the ordinary perturbative contributions to this series, and
finds that they are the same for $\theta=0$ and $\pi$, even though the
particles and $S$ matrices are completely different \cite{sausage}.
This is as it
must be: instantons and the $\theta$ term are a boundary effect
and hence cannot be seen in ordinary perturbation theory. One can also
identify the non-perturbative contributions to these series, and see
that they differ. Far away from the trivial fixed point,
the non-perturbative contributions dominate and cause
a non-trivial fixed point to appear when $\theta=\pi$.
The computation of the energy at zero temperature
in a background field was done for the $SU(N)/SO(N)$ and
$O(2P)/O(P)\times O(P)$ sigma models in \cite{metheta}.

In this paper I will compute the exact free energy at any temperature,
and thus compute the $c$ function. I will use a technique called the
thermodynamic Bethe ansatz (TBA), which I will describe in this section.

\subsection{The exact $S$ matrix}

An integrable field theory possesses an infinite number of conserved
currents and charges. The symmetries strongly constrain the dynamics,
but without making the system trivial. The constraints are why the
theory is ``solvable''. In this context, solvable means that some
quantities can be computed exactly.  These constraints imply that once
the particle spectrum is known, the exact $S$ matrix can be found.
Integrable models have the striking property that in a collision all
momenta are conserved individually, and that the $n$-body $S$ matrix
factorizes into a product of two-body ones. This two-body $S$ matrix
is completely elastic, meaning that the momenta and energy of the
particles are conserved individually, not just overall. Internal
quantum numbers can change in a collision, so the $S$ matrix is not
necessarily diagonal.  There are two possible ways of factorizing the
three-particle amplitude into two-particle ones; the requirement that
they give the same answer is the Yang-Baxter equation. There have been
hundreds of papers discussing how to solve this equation, so I will
not review this here. For a detailed discussion relevant to the sigma
models here, see e.g.\ \cite{ZZ,KT,OGW,Mackay}. Solutions arising in
the sigma models will be given below.

One of the useful characteristics of having particles in
representations of a Lie algebra is that their $S$ matrix can be
written in terms of projectors onto representations of this algebra.
The invariance of the $G/H$ sigma model under the Lie-group symmetry
$G$ requires that the $S$ matrices commute with all group
elements. The $S$ matrix can then be conveniently written in terms of
projection operators.  A projection operator ${\cal P}_a$ maps the
tensor product of two representations onto an irreducible
representation labelled by $a$. By definition, these operators satisfy
${\cal P}_a{\cal P}_b=\delta_{ab}{\cal P}_b$. Requiring invariance
under $G$ means that the $S$ matrix for a particle in the
representation $a$ with one in a representation $b$ means that the $S$
matrix is of the form
\begin{equation}
S^{ab}(\beta) = \sum_c f^{ab}_c (\beta) {\cal P}_c
\label{project}
\end{equation}
where $\beta\equiv\beta_a-\beta_b$ is the difference of the
rapidities, and the $f^{ab}_c$ are as of yet unknown functions. The
sum on the right-hand side is over all representations $c$ which
appear in the tensor product of $a$ and $b$; of course $\sum_c {\cal
P}_c=1$. In an integrable theory, the functions $f^{ab}_c(\beta)$ are
determined by requiring that the the $S$ matrix satisfy the
Yang-Baxter equation.

I define the prefactor $F^{ab}(\beta)$ to be the coefficient
$f^{ab}_c$ in (\ref{project}) where the highest weight of the
representation $c$ is the sum of the highest weights of the
representations $a$ and $b$. The Yang-Baxter equation does not give
this prefactor. To obtain it,
one needs to require that the $S$ matrix be unitary, and that it obey
crossing symmetry. With the standard assumption that the amplitude is
real for $\beta$ imaginary, the unitarity relation
$S^\dagger(\beta)S(\beta)=I$ implies $S(\beta)S(-\beta)=I$. The
latter is more useful because it is a functional relation which can be
continued throughout the complex $\beta$ plane.  Crossing symmetry is
familiar from field theory, where rotating Feynman diagrams by $90^o$
relates scattering of particles $a_i$ and $b_j$ to the scattering of
the antiparticle $\bar a_i$ with $b_j$.

Multiplying any $S$ matrix by function $F(\beta)$ which satisfies
$F(\beta)F(-\beta)=1$ and $F(i\pi-\beta)=F(\beta)$ will give an
$S$ matrix still obeying the Yang-Baxter equation, crossing and
unitarity (this is called the CDD ambiguity).  To determine
$F(\beta)$ uniquely, one ultimately needs to verify that the $S$
matrix is consistent with the bound-state structure,
and that it gives the correct $c$ function.

\subsection{Fusion}

In this paper, I derive the TBA equations for the sigma models by
utilizing fusion. Fusion is a method of finding new solutions of the
Yang-Baxter equation from known ones \cite{KRS}.  One starts with a
solution where the states are in some representation of a symmetry
algebra. Then one can find new solutions in other representations,
just as one takes tensor products of representations.  The usual place
fusion appears in the study of exact $S$ matrices is in what is called
the bootstrap (see e.g.\ \cite{OGW}). In many integrable models,
various particles can be thought of as bound states of other
particles.  The bootstrap procedure relates the $S$ matrices of bound
state to those of its constituents.  However, fusion is a more general
procedure than just the bootstrap.  It can be used to relate $S$
matrices of different models. This fact will prove very useful here,
because when the $S$ matrices are related, the TBA equations are
related as well. This observation enables the computation of the TBA
equation for integrable sigma models.

Formally speaking, fusion relies on the observation that at certain
values of $\beta$, the coefficients of some of the projectors in the
$S$ matrix vanishes. This means that some particles can be treated as
being composites: they are composed of ``constituent'' particles at
specific rapidities. I avoid calling the composite particles bound
states, because this implies that the composites and the constituents
are both particle states in the same theory. This is the not case in
general. For example, the only particles in the sine-Gordon model at
$\beta^2=8\pi$ particles are in the spin-$1/2$ representation of
$SU(2)$, while in the sphere sigma model, the only particles are in
the spin-$1$ representation of $SU(2)$. Fusion means that the $S$
matrices are related, even though the theories are different: the
spin-$1$ particles are composites of the spin-$1/2$ ones.

I will demonstrate fusion in theories with $SU(N)$ symmetry.  The
two-particle $S$ matrix for two particles in the $N$-dimensional
vector representations of $SU(N)$ contains two terms: one involving
the projector ${\cal P}_S$ onto the symmetric representation, the
other ${\cal P}_A$ onto the antisymmetric representation.  This is
because the tensor product of two symmetric representations in $SU(N)$
decomposes into the irreducible symmetric ($N(N+1)/2$ dimensional) and
antisymmetric representations ($N(N-1)/2$ dimensional):
 $$(N)\otimes (N) = (N(N-1)/2) \oplus (N(N+1)/2).$$  For $SU(2)$,
the antisymmetric representation is the singlet, so this statement
means that two spin $1/2$ representations tensored together is
the sum of the spin-0 and the spin-1 representations. The vector-vector
$S$ matrix for $SU(N)$ is determined by requiring that it satisfy the
Yang-Baxter equation.
It is
\begin{equation}
S^{VV}(\beta)=F^{VV}(\beta)
\left({\cal P}_S + \frac{\beta+2i\pi/N}{\beta-2i\pi/N}
\, {\cal P}_A \right).
\label{vvsmat}
\end{equation}
The function $F^{VV}(\beta)$ is the prefactor I defined above.
It must be consistent
with unitarity, crossing and the bootstrap.
A ``minimal'' solution of these constraints
means the $S$ matrix has no poles in the region
$0<Im(\beta) <\pi$. The minimal solution here is
\begin{equation}
F^{VV}_{min}(\beta)= \frac{\Gamma\left(1-\frac{\beta}{2\pi i}\right)
\Gamma\left(\frac{\beta}{2\pi i}+\frac{1}{N} \right)}
{\Gamma\left(1+\frac{\beta}{2\pi i}\right)
\Gamma\left(-\frac{\beta}{2\pi i}+\frac{1}{N}\right)}
\label{vvmin}
\end{equation}
For a given model, the prefactor $F^{ab}(\beta)$ may or may not be the
minimal solution. This prefactor is crucial to the physics, but
the fusion procedure is valid for any $F^{ab}(\beta)$.  

At $\beta=-2\pi i/N$, $S^{VV}$ in (\ref{vvsmat}) involves only the
projector onto the symmetric representation.  The fusion procedure
means that particles of rapidity $\beta_S$ in the symmetric
representation can be treated as being composed of two constituents in
the vector representation, of rapidities $\beta_S - i\pi /N$ and
$\beta_S + i\pi/N$.  The reason this works is described in \cite{KRS}.
The variable $\beta$ in the $S$ matrix is the difference of the
rapidities of the two particles, so when $\beta=2\pi i/N$, the
antisymmetric combination is effectively projected out. The Yang-Baxter
equation ensures that this projection survives any scattering.
In other words, if two vector particles are in the symmetric combination, 
they can scatter from other
particles and change state. However, if their rapidity difference
is $2\pi i/N$, the final state of these two particles will still be 
part of the symmetric representation.

Because particles in the symmetric representation are composed
of vector constituents, the $S$ matrices are related as well. The
$S$ matrix for scattering two particles in the symmetric representation
has three terms. In the language of weights \cite{Cahn}, the symmetric
representation has highest weight $2\mu_1$, and the tensor product is
$$ (2\mu_1) \otimes (2\mu_1) = (4\mu_1) \oplus (2\mu_1+\mu_2) \oplus (2\mu_2)$$
The $S$ matrix is
\begin{eqnarray}
S^{SS}(\beta)= F^{SS}(\beta)\left(
{\cal P}_{4\mu_1} + \frac{\beta+4\pi i/N}
{\beta-4\pi i/N}\,{\cal P}_{2\mu_1+\mu_2} + \frac{\beta+2\pi i/N}
{\beta-2\pi i/N}\,\,
\frac{\beta+4\pi i/N}{\beta-4\pi i/N}\,
{\cal P}_{2\mu_2}\right).
\label{sssmat}
\end{eqnarray}
The explicit form of the projection operators is given in
\cite{metheta}.  The minimal solution of the unitarity and crossing
constraints $F_{min}^{SS}(\beta)$ has no poles in the region
$0<Im\beta<\pi$, and is
\begin{equation}
F^{SS}_{min}(\beta)=\frac{\beta-2\pi i/N}{\beta+2\pi i/N}
\frac{\Gamma\left(1-\frac{\beta}{2\pi i}\right)
\Gamma\left(\frac{\beta}{2\pi i}+2\frac{1}{N}\right)}
{\Gamma\left(1+\frac{\beta}{2\pi i}\right)
\Gamma\left(-\frac{\beta}{2\pi i}+2\frac{1}{N}\right)}.
\label{ssmin}
\end{equation}
Note that $F^{SS}_{min}(\beta)$ differs from $F^{VV}_{min}(\beta + 2\pi i/N)
(F^{VV}_{min}(\beta))^2 F^{VV}_{min}(\beta - 2\pi i /N)$;
the prefactor does not automatically follow from the fusion procedure.

In cases where the composites are bound states of the constituents
(all are particles in the same theory), then the bootstrap procedure
relates the prefactors of composite scattering to those of constituent
scattering. However, the fusion does {\it not} make such a requirement
in general: the prefactor $F^{SS}(\beta)$ does not necessarily follow
from $F^{VV}(\beta)$. All the fusion procedure does is determine the
overall form of the $S$ matrix for the composite particles and ensure
that it obeys the Yang-Baxter equation.  Although one might expect
that $F^{SS}(\beta)= F^{VV}(\beta+2i\pi/N)(F^{VV}(\beta))^2
F^{VV}(\beta-2i\pi/N)$, I will show that below this is not true in
general here.  In another words, the CDD ambiguity may be resolved in
different ways in the constituent and composite theories.

\subsection{The free energy of an integrable theory}

Once the exact $S$ matrix is known, the exact free energy as a
function of mass, temperature, and magnetic field can be computed by
using the thermodynamic Bethe ansatz (TBA) \cite{YY,Ztba}.  This
enables one, for example, to compute thermodynamic quantities like the
susceptibility.  It also allows a very substantial check on any
assumption of integrability.  The reason is that at a critical point,
the free energy is known exactly -- it is related to the central
charge of the corresponding conformal field theory \cite{BCNA}. Thus
the free energy computed from the TBA must give this result in the
limit where the mass of the particles goes to zero, and the system is
at the unstable UV fixed point.

The TBA requires a relation
between the density of states of the particles to the actual
particle density. This relation is called the Bethe equation.
If the particles are free, this is trivial: the
density of states is independent of the particle density. If the
scattering is completely elastic and diagonal, this relation is easy
to derive. This is because a diagonal two-particle $S$
matrix is the boundary condition the phase shift in the wave function:
\begin{eqnarray}
\nonumber 
\psi(x_1,x_2) &&= e^{ip_1x_1 + ip_2 x_2} \qquad\qquad\qquad
\hbox{for } x_1\ll x_2\\ 
\psi(x_1,x_2) &&= e^{ip_1x_1 + ip_2 x_2}
S(p_1,p_2)\qquad\, \hbox{for } x_1\gg x_2
\label{phaseshift}
\end{eqnarray}
In a state of ${\cal N}$ particles, the Bethe equation follows by
requiring that one-dimensional space of length $L$ be periodic, and
that the wavefunction be invariant under sending any of the
coordinates $x_i\to x_i+L$. First consider the case where there is
only one kind of particle in the spectrum, with two-particle $S$
matrix $S(\beta_1-\beta_2)$.  The requirement of periodicity of the
wavefunction $\psi(x_1,x_2,\dots x_{\cal N})$ yields the relations
\begin{equation}
e^{im\sinh\theta_iL}
\prod_{j=1}^{\cal N} S(\beta_i-\beta_j)=1
\label{quant}
\end{equation}
One can think of this intuitively as bringing the particle around the
world through the other particles; one obtains a product of
two-particle $S$-matrix elements because the scattering is
factorizable. 
This is the generalization of the free-particle momentum quantization condition
$p=2n\pi/L$.

The Bethe equation is written in terms of the density of states
$P(\beta)$ and the density of rapidities $\rho(\beta)$.
The former is defined so that the number of allowed states with
rapidities between $\beta$ and $\beta+d\beta$ is $P(\beta)d\beta$,
while the number of states actually occupied in this interval
is $\rho(\beta)d\beta$. The quantization condition relates the two.
Taking the derivative of the log of (\ref{quant}) yields
\begin{equation}
2\pi P(\beta) =mL\cosh\beta+
\int_{-\infty}^\infty d\beta'\,\Phi(\beta-\beta')\rho(\beta')
\label{bethe}
\end{equation}
where
$\Phi(\beta)={1\over i}{d\over d\beta}\ln S(\beta).$
This is easily generalized to the situation where there is more than
one particle in the spectrum, as long as the scattering is diagonal.
Let $S_{ab}$ be the $S$ matrix element for scattering a particle of
type $a$ from one of type $b$.
Defining densities $P_a$ and $\rho_a$ for each type of particle,
the Bethe equations are
\begin{equation}
2\pi P_a(\beta) =m_aL\cosh\beta+
\sum_b \int_{-\infty}^\infty d\beta'\,\Phi_{ab}(\beta-\beta')\rho_b(\beta') .
\label{bethe2}
\end{equation}
where $$\Phi_{ab} = {1\over i}{d\over d\beta}\ln S_{ab}(\beta)$$

Once the Bethe equations are known, the TBA equations and hence the
free energy can be derived. This is done by minimizing the free energy,
using (\ref{bethe2}) as a constraint. The result is most conveniently written
in terms of the ``dressed particle energies'' $\epsilon_a(\beta)$,
defined by
\begin{equation}
\frac{\rho_a(\beta)}{P_a(\beta)} =
\frac{1}{1+e^{-\epsilon_a(\beta)/T}}.
\label{epdef}
\end{equation}
For simplicity, I have set all
chemical potentials and background fields to be zero.
The resulting TBA equations are \cite{YY,Ztba}
\begin{equation}
\epsilon_a(\beta) = {m_a}\cosh\beta - 
\sum_b
T \int_{-\infty}^{\infty} \frac{d\beta'}{2\pi} \Phi_{ab}(\beta-\beta') 
\ln\left( 1 + e^{-\epsilon_b(\beta)/T}\right)
\label{tba}
\end{equation}
For free particles, $\Phi_{ab} =0$ and the $\epsilon_a$ just reduce to
the particle energies.  The form of the TBA equations reflects the
fact that in all integrable particle theories of this type, it is
either proven or assumed that the particles fill levels like fermions:
at most one particle in a level.
The free energy per unit length $F$
is given in terms of these dressed energies $\epsilon_a$. It is
\begin{equation}
F (m,T)=
 - T \sum_a m_a \int_{-\infty}^{\infty} \frac{d\beta}{2\pi} \cosh\beta
\ln\left( 1 + e^{-\epsilon_a(\beta)/T}\right)
\label{free}
\end{equation}
In the IR limit $m_a\to\infty$, the gas of
particles becomes dilute, and interactions can be neglected.
The free energy becomes 
\begin{equation}
\lim_{m_a\to\infty} F (m,T) = 
-T \sum_a m_a\int_{-\infty}^{\infty} \frac{d\beta}{2\pi} \cosh\beta
e^{-m_a\cosh(\beta)/T}
\label{freeir}
\end{equation}
This integral can be done, yielding a Bessel function.

Calculating the free energy using the TBA allows an extremely
non-trivial check on the exact $S$ matrix.  In the limit of all masses
going to zero, the theorem of \cite{BCNA} says that the free energy
per unit length must behave as
\begin{equation}
\lim_{m_a\to 0} F = -\frac{\pi T^2}{6} c_{UV}
\label{freeuv}
\end{equation}
where $c_{UV}$ is the central charge of the conformal field theory
describing this UV limit. The number $c_{UV}$ can usually be
calculated analytically from the TBA, because in this limit the free
energy can be expressed as a sum of dilogarithms \cite{KR}.  The
$c_{UV}$ computed from the TBA must of course match the $c_{UV}$ from
the field theory. This provides an extremely non-trivial check not
only of the $S$ matrix, but of whether the entire spectrum is
known. All particles contribute to the free energy, so if some piece
of the spectrum is missing or if an incorrect particle is included, the
correct $c_{UV}$ will not be obtained.

The TBA computation is much trickier if the scattering between
particles is non-diagonal, as is the situation for the models of
interest here.  The Bethe equation is much harder to derive,
because as one particle is going around the periodic world,
it can change state as it scatters though the other particles.
This requires introducing the ``transfer matrix''
${\cal T}$ for
bringing the a given particle through the others; since the
scattering is not diagonal, the final state is not necessarily the
same as the initial.
To define ${\cal T}$ explicitly,
I first introduce the scattering matrix  ${\cal T}_{ab}(\beta)$
for bringing
a particle of type $a$ and rapidity $\beta$
through ${\cal N}$ particles and ending up
with a particle of type $b$. 
Thus the different ${\cal T}_{ab}$ 
make up a set of $s^2$ $s^{{\cal N}}\times
s^{{\cal N}}$ matrices, where $s$ is the number of different types
of particles.
The scattering is completely elastic, so the rapidities do not
change even though the scattering is not diagonal. This means
${\cal T}_{ab}(\beta)$
depends on the rapidities $\beta_1\dots \beta_{\cal N}$ as well as $\beta$.
Let $S_{ab\to cd}(\beta_1-\beta_2)$ 
be the two-particle $S$ matrix element for
scattering an initial state $a(\beta_1) b(\beta_2)$
and ending with a final state of $c(\beta_2)d(\beta_1)$.
Then the components of
${\cal T}_{ab}$ can be written in terms of the $S$ matrix elements as
$$({\cal T}_{ab}(\beta|\beta_1\dots\beta_{\cal N}))_{c_1c_2\dots c_{\cal N}}
^{d_1d_2\dots d_{\cal N}} \equiv
\sum 
S_{ac_{1}\to d_{1}f_{1}}(\beta -\beta_{1})
S_{f_{1}c_{2}\to d_{2}f_{2}}(\beta-\beta_{2})\dots
S_{f_{\cal N}c_{\cal N}\to d_{\cal N}b}(\beta-\beta_{\cal N})$$
where the sum is over the intermediate states 
$f_1=1\dots s$, $f_2 = 1\dots s$, \dots,
$f_{\cal N}=1\dots s$. The matrix ${\cal T}$ follows by exploiting the
fact that all the $S$ matrices of interest 
at zero relative rapidity just permute the
colliding particles.  In other words, 
$S_{ab\to cd}(0)=-\delta_{ac}\delta_{bd}$. Thus setting
$\beta=\beta_\alpha$ effectively turns the $\alpha^{\hbox{th}}$
particle so that it scatters through all the others.  This is
precisely what is needed for the TBA. To put periodic boundary
conditions on the system, one sums ${\cal T}_{aa}$ over all $a$. The
result is that
\begin{equation}
{\cal T}(\beta_\alpha| \beta_{1},\dots \beta_{\cal N})
\equiv \sum _{a}{\cal T}_{aa}
(\beta=\beta_\alpha|\beta_{\alpha+1},\dots \beta_{\cal N},\beta_1,\dots,
\beta_{\alpha-1}).
\label{Tmat}
\end{equation}
This is a
$s^{{\cal N}-1}\times
s^{{\cal N}-1}$ matrix.

The TBA requires finding the 
eigenvalues 
$\Lambda(\beta_\alpha| \beta_{1},\dots \beta_{\cal N})$ of ${\cal T}$. 
The crucial effect of the $S$ matrix satisfying
the Yang-Baxter relation is that the ${\cal T}(\beta)$
commute for different $\beta$. This ensures that ${\cal T}(\beta)$
can be
simultaneously diagonalized for all $\beta$ by a $\beta$-independent
set of eigenvectors; only the eigenvalues depend on $\beta$.
The
quantization condition (\ref{quant}) is generalized to
\begin{equation}
e^{im_\alpha\sinh \beta _\alpha L}
{\Lambda}(\beta _\alpha |\beta _{1},\dots \beta_{\cal N})=1
\label{quant2}
\end{equation}
This must hold for all particles $\alpha=1\dots {\cal N}$.
In the limit of large ${\cal N}$, $\Lambda$ depends
on the particle densities instead of the individual rapidities.
Henceforth I will just write $\Lambda(\beta)$.
For the cases
of interest here, finding the eigenvalues $\Lambda(\beta)$
is quite difficult, but has been
done in \cite{BRi,BRii,BRiii}. The Bethe equations are still of the form
(\ref{bethe2}), and the TBA equations are still of the form
(\ref{tba}). However, extra particles, known as ``pseudoparticles''
or ``magnons'', enter the equations. These particles appear in the
equations just as if they were a particle species, but with $m_a=0$.
I will give examples of the explicit form of these equations below.

The transfer matrix has very nice properties under fusion, because
the fused $S$ matrices are products of the constituent $S$ matrices.
The case of most interest here is when particles in the
representation with highest weight $\mu_a$ are fused to give particles 
in the representation $2\mu_a$.  Then
the transfer matrices for ${\cal N}/2$ fused particles is related
to the product of transfer matrices for ${\cal N}$ constituents. The
reason it is a product is because both constituents must be brought
around the world in the fused transfer matrix. The precise relation is
\begin{eqnarray}
\nonumber
{\cal T}^{2\mu_a}
(\beta_\alpha| \beta_{1},\dots \beta_{{\cal N}/2})&= & C(\beta_\alpha)
{\cal T}^{\mu_a}(\beta_\alpha+\eta| 
\beta_{1}+\eta,\beta_{1}-\eta,\dots 
\beta_{{\cal N}/2}+\eta,\beta_{{\cal N}/2}-\eta) \times\\&&
\ {\cal T}^{\mu_a}(\beta_\alpha-\eta| \beta_{1}+\eta,
\beta_{1}-\eta,\dots 
\beta_{{\cal N}/2}+\eta,\beta_{{\cal N}/2}-\eta)
\label{tfused}
\end{eqnarray}
The rapidity difference of the constituents is $2\eta$.
The reason for the extra factor $C(\beta)$
is that the prefactors of the $S$ matrices
need not satisfy the exact fusion relation, as discussed above.  
This constant of proportionality is
$$C(\beta) = \prod_{\alpha =1}^{{\cal N}/2} 
\frac{F^{2\mu_a\, \nu_\alpha}(\beta-\beta_\alpha)}
{F^{\mu_a \nu_\alpha}(\beta-\beta_{\alpha}+\eta)
F^{\mu_a \nu_\alpha}(\beta-\beta_{\alpha}-\eta)}
$$
where the particle with rapidity $\beta_\alpha$ is in the representation
$\nu_\alpha$.
Given this relation between transfer matrices, the eigenvalues 
obey the relation
\begin{equation}
\Lambda^{2\mu_a}(\beta) = C(\beta)\Lambda^{\mu_a}(\beta+\eta)
\Lambda^{\mu_a}(\beta-\eta) .
\label{evfused}
\end{equation}

\section{Massive sigma models}

In this section I will derive the TBA equations
for a variety of massive sigma models. I start with the
sphere sigma model, before going on to the more complicated
cases.

\subsection{The sphere sigma model}

One of the best-known sigma models is the sphere sigma model, where
the field takes values on a two-sphere. In the $G/H$ language
I have been using, this corresponds to $G=SU(2)$ or $G=SO(3)$, and
$H=U(1)$ or $H=SO(2)$. The TBA equations were derived originally by
taking the limit of certain integrable fermion models
\cite{WiegO3,Tsvel}, and conjectured on different grounds in
\cite{FZ}. I will rederive the TBA equations here directly from
the $S$ matrix, because this is the method which generalizes most
simply to the more general sigma models of interest.

In a two-dimensional $G/H$ sigma model, the global symmetry group is
$G$. Therefore
the symmetry group of the sphere sigma model is $G=SO(3)$: the symmetry
corresponds to rotations of the sphere.
The particles of
this model were shown long ago to be in the spin-$1$ representation
of $SO(3)$ \cite{ZZ}. Their $S$ matrix was derived by solving the
Yang-Baxter equation directly, and is given by (\ref{sssmat})
with $N=2$ and
\begin{equation}
F^{SS}_{N=2}(\beta)=\frac{\beta-i\pi}{\beta + i\pi}
\label{presphere}
\end{equation}

Since this $S$ matrix is non-diagonal, one needs to diagonalize the
transfer matrix as described in the last section. The way to do this
is to first solve the problem for particles in the spin-$1/2$
representation of $SU(2)$, and then use fusion to find the answer for
spin $1$. For particles in the spin-$1/2$ representation of $SU(2)$,
the two-particle $S$ matrix is given by (\ref{vvsmat}) with $N=2$.
This $S$ matrix is four-by-four, since there are just two different
kinds of particles (spin up and down).
The choice
$$F^{VV}_{N=2} = F^{VV}_{min}$$ gives the $S$ matrix of the sine-Gordon
model at the coupling $\beta^2=8\pi$ in the usual conventions. At this
coupling, the dimension of the $\cos\beta\phi$ perturbation is two, so
that it is marginally relevant; the $U(1)$ symmetry of the sine-Gordon
model is enhanced to $SU(2)$. Another name for this model is
the $SU(2)$ Gross-Neveu model.

For particles in the spin-$1/2$ representation of $SU(2)$, the Bethe
equations were derived 70 years ago, in the original paper by Bethe
himself \cite{Bethe}. The reason is that the transfer matrix for the
spin-$1/2$ representation of $SU(2)$ as defined in (\ref{Tmat})
precisely corresponds to the transfer matrix of the Heisenberg spin
chain.  In the limit of large number of particles ${\cal N}$,
the eigenvalues of
the transfer matrix follow by adopting the ``string hypothesis''.
This means that the eigenvalues $\Lambda(\beta)$ of the
transfer matrix defined in (\ref{Tmat}) are expressed in terms of
densities $\widetilde{\rho}_k(\beta)$, with $k=1\dots \infty$. 
These are the
pseudoparticles discussed above: they enter the TBA equations as if
they were real particles with no mass term. (I have somewhat
abused the conventional notation: most authors would not
use the $\tilde{}$ here, but it makes subsequent relations less confusing.)
The other density entering
the equations is the density of particles $\rho_0(\beta)$.  This is
the total particle density, with contributions of both spin up and
spin down particles. 

Bethe's result for the eigenvalues is
\begin{equation}
\frac{d}{d\beta} \ln\Lambda(\beta) = 
Y^{(2)}*\rho_0(\beta) + \sum_{j=1}^\infty
\sigma^{(\infty)}_{j} * \widetilde{\rho}_j(\beta)
\label{evsG}
\end{equation}
where 
convolution integrals are defined as
$$f*g(\beta) = \int_{-\infty}^\infty d\beta' f(\beta-\beta')
g(\beta).$$ The kernels are given explicitly in the Appendix. The kernel
$Y^{(N)}$ comes from the prefactor of the $S$ matrix. This only
affects the coupling to the total particle density, and not the
pseudoparticles, because it contributes an overall factor
$\prod_{\alpha=1}^{\cal N} F^{VV}(\beta-\beta_{\alpha})$ 
to the transfer
matrix.
Now I can write down the first of the Bethe equations, by taking the
derivative of the log of (\ref{quant2}). 
This gives
\begin{equation}
2\pi P_0 (\beta) = {m}\cosh\beta + 
Y^{(2)}*\rho_0(\beta) - \sum_{j=1}^\infty
\sigma^{(\infty)}_{j} * \widetilde\rho_j(\beta).
\label{bet0}
\end{equation}
where $m$ is the mass of the particles.
$P_0$ is the total density of states for the particles.
The other Bethe equations relate the densities of states for
the pseudoparticles to particle and pseudoparticle densities.
They are 
\begin{equation}
2\pi {\rho}_j(\beta) = 
\sigma_j^{(\infty)} * \rho_0(\beta) - \sum_{l=1}^\infty
A_{jl}^{(\infty)} * \widetilde{\rho}_l(\beta)
\label{betj}
\end{equation}
where the density of string states $P_j$ is
$$P_j = \widetilde{\rho}_j + \rho_j$$
Note that all the Bethe equations are of the form (\ref{bethe2}),
with no mass term for the pseudoparticles.

Using identities in the appendix, 
all the Bethe equations (including that for $P_0$) 
can be written in the compact form
\begin{equation}
2\pi {P}_j(\beta) = \delta_{j0} m \cosh\beta + 
\sum_{l=0}^\infty I^{(\infty)}_{jl}\int_{-\infty}^{\infty}
d\beta' \frac{1}{\cosh(\beta-\beta')} \rho_l(\beta')
\label{betall}
\end{equation}
Here the indices $j$ and $l$ in the incidence matrix
$I^{(\infty)}_{jl} = \delta_{j,l+1} + \delta_{j,l-1}$
run from $0,1,\dots,\infty$. Note that the right-hand-side
involves the hole densities, not the particle densities.
This Bethe equation is conveniently represented by the diagram
in figure 1.
With these equations, it follows from the standard TBA calculation that
the TBA equations (\ref{tba},\ref{free}) hold, with
$$\Phi_{jl}(\beta) = 
\frac{I^{(\infty)}_{jl}}{\cosh(\beta)}$$
and
$$m_{j} =\delta_{j0}m\cosh\beta.$$
These equations were first derived in the context of the sine-Gordon model
at $\beta^2\to 8\pi$ in \cite{fowler}. 
One can easily check that the free energy has the correct properties.
In the UV limit $m/T\to 0$, one obtains the correct central charge $c_{UV}=1$
from (\ref{freeuv}).
This follows from a now-standard analysis, involving expressing
the free energy as a sum of dilogarithms (see e.g.\ \cite{KR,Ztba,FS}).
In the IR limit, the generalization of (\ref{freeir}) to the
case with pseudoparticles is
$$F = 
 mT \left(1+e^{-\epsilon_1(\infty)}\right)^{1/2}
\int_{-\infty}^{\infty} \frac{d\beta}{2\pi} \cosh\beta
e^{-m\cosh(\beta)/T}
$$
For particles with $m_j\ne 0$, $e^{-\epsilon_j(\infty)}$
vanishes. However, the pseudoparticles have no mass term, and here one
finds that $e^{-\epsilon_j(\infty)} = (j+1)^2 -1$ for $j\ge 1$. This
means that the free energy in the IR limit is that of $2$ types
of particles of
mass $m$, as it must be.
\begin{figure}
\centerline{\epsfxsize=4.0in\epsffile{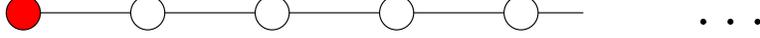}}
\bigskip\bigskip
\caption{The incidence diagram for the $SU(2)$ Gross-Neveu model (the 
sine-Gordon at $\beta^2\to 8\pi$). The circles represent the functions
$\epsilon_a$; the filled node represents the fact that the equation
for $\epsilon_0$ has a mass term. The line represents the coupling
between the functions in the TBA equations.}
\end{figure}

\bigskip

It is now simple to get the $S$ matrices and TBA for the
sphere sigma model by using fusion. The fusion procedure says
that the spin-$1$ particles in
the sphere sigma model can be viewed as having the spin-$1/2$
particles as constituents. As explained above, a spin-$1$
particle (in a representation with highest weight $2\mu_1$) 
is composed of a pair spin-$1/2$ particles (each in a representation
with highest weight $\mu_1$) 
with
rapidities $\beta_i + i\pi/2$
and $\beta_i - i\pi/2$.
The transfer matrix for ${\cal N}/2$
spin-$1$
particles 
is related to that for the ${\cal N}$ spin-$1/2$ particles 
by the relation (\ref{tfused}) with $\eta=i\pi /2$.
Because the two transfer matrices are related in this way, the Bethe
equations for the sphere sigma model follow from those above after a
few modifications.
The eigenvalue of the sphere sigma model transfer matrix follows
from the spin-$1/2$ eigenvalue (\ref{evsG}), and
the fusion equation (\ref{evfused}).
It is
\begin{equation}
\frac{d}{d\beta} \ln\Lambda^{sphere}(\beta) = 
Z^{(2)}*\rho_0(\beta) + \sum_{j=1}^\infty
\tau^{(\infty)}_{j} * \rho_j(\beta)
\label{evsphere}
\end{equation}
where 
$$ \tau^{(\infty)}_j (\beta) = 
\sigma^{(\infty)}_j (\beta+i\pi/N) +
\sigma^{(\infty)}_j (\beta-i\pi/N)$$
with $N=2$ here.
The first term in (\ref{evsphere}) arises from the prefactor of the
sphere $S$ matrix (\ref{presphere}), with
$$Z^{(2)} = -i\frac{\partial}{\partial\beta} \ln\, F^{SS}_{N=2} = 
\frac{2\pi}{\beta^2 + \pi^2}.$$
The explicit expressions for $\tau^{(s)}$ and $Z^{(N)}$
are given in 
(\ref{tauj}) and (\ref{eta}) in the appendix.
Using this expression for the eigenvalue in (\ref{quant2}) gives
\begin{equation}
2\pi P_0 (\beta) = {m}\cosh\beta + 
Z^{(2)}*\rho_0(\beta) - \sum_{j=1}^\infty
\tau^{(\infty)}_{j} * \rho_j(\beta).
\label{bets0}
\end{equation}
The Bethe equations for the densities
of states of the pseudoparticles (\ref{betj})
are modified because  the real particles come in pairs
with rapidities $\beta\pm i\pi/2$. Thus for the sphere sigma model
\begin{equation}
2\pi {\rho}_j(\beta) = 
\tau_j^{(\infty)} * \rho_0(\beta) - \sum_{l=1}^\infty
A_{jl}^{(\infty)} * \widetilde\rho_l(\beta)
\label{betsj}
\end{equation}
for $j\ge 1$.

By using the identities in the appendix, the Bethe equations
(\ref{betsj}, \ref{bets0}) can
be put in the unified form 
\begin{equation}
2\pi {P}_j(\beta) = \delta_{j0} m \cosh\beta + 
\sum_{l=0}^\infty {\cal I}^{(\infty)}_{jl}\int_{-\infty}^{\infty}
d\beta' \frac{1}{\cosh(\beta)} \rho_l(\beta').
\label{betsall}
\end{equation}
The indices $j$ and $l$ here run from $0\dots\infty$.
Above, the incidence matrix $I^{(s)}$ was associated
with $SU(s)$. Here, the incidence matrix ${\cal I}^{(s)}$ is 
associated with $O(2s)$: ${\cal I}^{(s)}_{jl}= 2\delta_{jl}-
C^{O(2s)}_{jl}$, where $C^{O(2s)}$ is the Cartan matrix for
$O(2s)$. Explicitly,
\begin{equation}
{\cal I}^{(\infty)}_{jl} = \delta_{j,l+1} + \delta_{j,l-1}
+\delta_{j,2}\delta_{l,0} + \delta_{j,0}\delta_{l,2}
-\delta_{j,1}\delta_{l,0} - \delta_{j,0}\delta_{l,1}
\label{incSO}
\end{equation}
This Bethe equation is conveniently represented by the diagram
in figure 2. 
\begin{figure}
\centerline{\epsfxsize=4.0in\epsffile{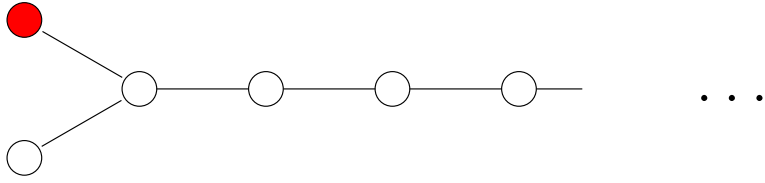}}
\bigskip\bigskip
\caption{The incidence diagram for the sphere sigma model}
\end{figure}

With these equations, it follows from the standard TBA calculation that
the TBA equations (\ref{tba},\ref{free}) hold, with
$$\Phi_{jl}(\beta) = 
\frac{{\cal I}^{(\infty)}_{jl}}{\cosh(\beta)}$$
and
$$m_{j} =\delta_{j0}m\cosh\beta.$$
One can easily check that the free energy has the correct properties
\cite{FZ}.
In the UV limit $m/T\to 0$, one obtains the correct central charge $c_{UV}=2$
by the standard dilogarithm analysis.
In the IR limit, one finds that
$$F = 
 mT \left(1+e^{-\epsilon_2(\infty)}\right)^{1/2}
\int_{-\infty}^{\infty} \frac{d\beta}{2\pi} \cosh\beta
e^{-m\cosh(\beta)/T}.
$$
As with the spin-$1/2$ system, the functions obey
$e^{-\epsilon_j(\infty)} = (j+1)^2 -1$ for $j\ge 1$. This
means that the free energy in the IR limit is that of $3$ types of
particles of
mass $m$, the spin-$1$ triplet.

\subsection{$SU(N)$ Gross-Neveu models}

To find the sigma model free energy,
it is best to first perform the analysis for the vector
particles and then use fusion.
The appropriate field theory with particles in the vector
representation of $SU(N)$ is the $SU(N)$ Gross-Neveu model 
(also sometimes called the chiral Gross-Neveu model) \cite{natan,sunGN}. Its
similarities and differences with the sigma model were discussed at
length in \cite{metheta}. The vector particles in the $SU(N)$ Gross-Neveu
model have the $S$ matrix (\ref{vvsmat}). The prefactor
$F_{GN}^{VV}(\beta)$ is not the minimal one given in
(\ref{vvmin}). It is instead
$$F_{GN}^{VV}(\beta)= F^{VV}_{min}(\beta) X(\beta)$$
where
\begin{equation}
X(\beta) = \frac{\sinh\left(\frac{1}{2}(\beta + 2\pi
i/N)\right)} {\sinh\left(\frac{1}{2}(\beta - 2\pi
i/N)\right)}.
\label{forX}
\end{equation}
Note that $X=1$ for $N=2$, so the sine-Gordon model at $\beta^2\to 8\pi$ is
indeed the $SU(2)$ Gross-Neveu model.

The pole at $\beta=2\pi i/N$ in this factor
$X(\beta)$ means that for $N>2$, the vector particles in the
Gross-Neveu model have bound
states in the antisymmetric representation. Upon completing
the bootstrap procedure, one finds that the
model has bound states in all the antisymmetric representations with
$a$ indices, $a=1\dots N-1$. 
These are called the fundamental representations, and they have
highest weight $\mu_a$. The particles
can be expressed as bound states of $a$ particles in the vector representation.
These have mass
$$m_a = m \sinh\left(\frac{\pi a}{N}\right)$$ 
The representation with
highest weight $\mu_{N-a}$ is the conjugate of the representation
$\mu_a$, because of the invariant $\epsilon$ tensor. For example, the
$\overline{N}$ representation has highest weight $\mu_{N-1}$ and mass
$m_{N-1} = m_1$.  The bootstrap procedure
gives the $S$ matrices for all scattering of these particles.
The scattering
is not diagonal, but it is diagonal in the representation labels.
When a particle in representation $a$
with rapidity $\beta_\alpha$
scatters, the final particle with
rapidity $\beta_\alpha$ must be in some state in same representation $a$. 
This means that the two-particle $S$ matrix
prefactors can be labelled by $F^{ab}$.
The vector-vector prefactor $F^{VV}\equiv F^{11}$ in this new notation.
The explicit prefactor $F^{ab}_{GN}$
is necessary for the calculation, and is given in
(\ref{gamm2}) in the appendix.

Computing the Bethe equations for the $SU(N)$ Gross-Neveu models
looks extremely difficult or impossible. Remarkably, the computation
has already been done in \cite{BRii,BRiii} by using fusion. Here the
Bethe equations are found for any simply-laced Lie algebra $G$, when
the particles are in any representations with highest weight $m\mu_i$
where $\mu_i$ is a fundamental weight of $G$, and $m$ is an
integer. This work was generalized to non-simply-laced groups in
\cite{KNS}.  The fusion procedure gives functional relations like
(\ref{tfused}) for all the ${\cal T}^{a}(\beta _\alpha|\beta
_{1},\dots ,\beta _{\cal N})$ \cite{BRii}. The label $a$ here
indicates that the $\alpha^{\hbox{th}}$ particle is in the
representation with highest weight $\mu_a$. These functional relations
relate various ${\cal T}^{a}$.  The prefactors $F^{ab}(\beta)$ need to
be computed, but the explicit $S$ matrix is not needed: all the
relevant physics is contained in the representation theory and in the
fusion.  {}From the functional relations and a few mild analyticity
assumptions, the eigenvalues of ${\cal T}^{a}$ and the Bethe equations
can be derived in the limit of a large number of particles.

The Bethe equations for the general case require the
introduction of pseudoparticle densities and densities of states into
the Bethe equation (\ref{bethe2}).  Here the
pseudoparticle densities $\widetilde{\rho}_{a,j}$ 
and densities of states $P_{a,
j}(\beta)$ are labelled by two indices.  (In the literature,
this is usually called a nested Bethe ansatz.)
The index $a$ runs from $1$
to $N-1$ for $SU(N)$. For the $N=2$ case treated above, this index
takes only one value can be suppressed.  The index $j$ is the same
index as before, running from $1,\dots\infty$ for the pseudoparticles.
The functions $\rho_{a,0}$ and $P_{a,0}$ are defined respectively
as the density and density of states for all the particles 
in the representation $\mu_a$. It is consistent
to define separate densities for each representation, because
the particles cannot change representation when scattering.
For all values of $a$ and $j$,
$P_{a,j}=\rho_{a,j} + \widetilde\rho_{a,j}$.

The computation of the TBA equations directly from the $SU(N)$ Gross-Neveu
model $S$ matrix was done in \cite{Hollo}.
The eigenvalues of the transfer matrix $T^{a}$ are \cite{BRii,BRiii}
\begin{equation}
\frac{d}{d\beta} \ln\Lambda_{GN}^a(\beta) = 
\sum_{b=1}^{N-1} Y^{(N)}_{ab}*\rho_{b,0}(\beta) + 
 \sum_{j=1}^\infty
\sigma^{(\infty)}_{j} * \widetilde{\rho}_{a,j}(\beta)
\label{evGN}
\end{equation}
where the kernels are given explicitly in the Appendix. The kernel
$Y^{(N)}_{ab}$ comes from the prefactor $F^{ab}$
of the $S$ matrix. It couples the density of states of
real particles in
representation $a$ to the density of particles in representation
$b$. The first of the Bethe equations follows from (\ref{quant2}),
and is
\begin{equation}
2\pi P_{a,0} (\beta) = {m_a}\cosh\beta + 
\sum_{b=1}^{N-1}Y^{(N)}_{ab}*\rho_{b,0}(\beta) - 
\sum_{j=1}^\infty
\sigma^{(\infty)}_{j} * \widetilde{\rho}_{a,j}(\beta).
\label{betGN}
\end{equation}
The other Bethe equations relate the densities of states for
the pseudoparticles to particle and pseudoparticle densities.
They follow from \cite{BRii,BRiii} as well, and are
\begin{equation}
2\pi {\rho}_{a,j}(\beta) = 
\sigma_j^{(\infty)} * \rho_{a,0} - \sum_{b=1}^{N-1}\sum_{l=1}^\infty 
A_{jl}^{(\infty)} * K_{ab}^{(N)} * \widetilde\rho_{b,l}(\beta)
\label{betGNj}
\end{equation}
where 
$P_{a,j} = \widetilde\rho_{a,j} + \rho_{a,j}$. Explicit expressions
for these kernels are given in the Appendix.
Note how all these equations reduce to those in the last subsection
by setting $N=2$.

By using the fact that $A$ and $K$ are inverses, and the identities
in the appendix, 
all the Bethe equations (\ref{betGN},\ref{betGNj})
can be written in the combined form \cite{Hollo}
\begin{equation}
2\pi \widetilde{\rho}_{a,j}(\beta) = \delta_{j0} m_a \cosh\beta -
\sum_{b=1}^{N-1}\sum_{l=0}^\infty 
{K^{(\infty)}_{jl}}
*A_{ab}^{N} * \rho_{b,l} (\beta)
\label{betGNall}
\end{equation}
Here the indices $j$ and $l$ 
run from $0,1,\dots,\infty$. 
With these densities, the dressed energies $\epsilon_{a,j}(\beta)$
are defined as in (\ref{epdef}). 
It follows from the standard TBA calculation that
the TBA equations (\ref{tba},\ref{free}) hold, with
$$\Phi_{ab,jl}(\beta) = \delta_{jl}\delta_{ab}\delta(\beta)
-K^{(\infty)}_{jl}*A_{ab}^{(N)}(\beta)$$
and
$$m_{aj} =\delta_{j0}m_a\cosh\beta.$$
The TBA equations can be rewritten in a much more elegant form
by using the fact that $A$ and $K$ are inverses,
and the simple relation between $K$ and the incidence
matrix
\begin{equation}
I^{(N)}_{jl}=\delta_{j,l-1} + \delta_{j,l+1}\qquad\qquad j,l=1\dots N-1
\label{inc}
\end{equation}
The result is
\begin{eqnarray}
\nonumber
\epsilon_{a,j} (\beta) &=& 
T\sum_{b=1}^{N-1}
I^{(N)}_{ab}\int_{-\infty}^\infty \frac{d\beta'}{2\pi}
\frac{N}{2\cosh(N(\beta-\beta')/2)}
\ln\left(1+e^{\epsilon_{b,j}(\beta')}\right)\\
&&\qquad- T\sum_{l=0}^{\infty}
I^{(\infty)}_{jl}\int_{-\infty}^\infty \frac{d\beta'}{2\pi}
\frac{N}{2\cosh(N(\beta-\beta')/2)}
\ln\left(1+e^{-\epsilon_{a,l}(\beta')}\right)
\label{tbaGN}
\end{eqnarray}
This is a substantial simplification because 
the equation for $\epsilon_{a,j}$ only involves ``adjacent'' functions
$\epsilon_{a,j\pm 1}$ and $\epsilon_{a\pm 1,j}$.
These equations are displayed schematically in figure 3.  
The dashed
and unbroken lines account for the different minus signs in
(\ref{tbaGN}).
Note that the masses do not appear in rewritten TBA equations
(\ref{tbaGN}), although they appear
in the original ones. When using the form (\ref{tbaGN}), the asymptotic
conditions
$$\epsilon_{a,0}(\beta\to\infty) \rightarrow m_a\cosh\beta.$$
must be imposed.

\begin{figure}
\centerline{\epsfxsize=2.5in\epsffile{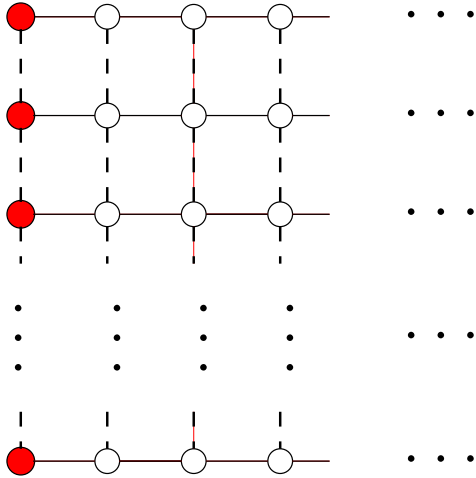}}
\bigskip\bigskip
\caption{The incidence diagram for the $SU(N)$ Gross-Neveu model. There
are $N-1$ rows and an infinite number of columns.}
\end{figure}

This free energy of the $SU(N)$ Gross-Neveu model
has the correct properties.  In the UV limit $m/T\to 0$, one obtains
the correct central charge $c_{UV}=N-1$  from the dilogarithm
analysis.  In the IR limit, one finds that each representation
contributes one term to the free energy, with the correct multiplicity
(e.g.\ $N$ for the vector representation $a=1$, $N(N-1)/2$ for the
antisymmetric representation $a=2$).

\subsection{$SU(N)/SO(N)$ sigma models}

Here I find the TBA equations for the $SU(N)/SO(N)$ sigma model,
generalizing the analysis for the sphere sigma model, which corresponds
to $N=2$. The TBA analysis is related to that for
$SU(N)$ Gross-Neveu models via fusion for all $N$.

The $SU(N)/SO(N)$ sigma models have a Lagrangian description 
(\ref{action}) in terms
of a a symmetric and unitary matrix field.  The particles of the sigma
model are in all representations with highest weight $2\mu_a$,
$a=1\dots N-1$ \cite{metheta}. The representation
with highest weight $2\mu_1$ is the symmetric representation. The 
two-particle $S$ with both particles in the symmetric representation
is given by (\ref{sssmat}) with prefactor \cite{metheta}
$$F^{SS}(\beta) = X(\beta) F^{SS}_{min}(\beta)$$
where the minimal factor is given in (\ref{ssmin}), and $X(\beta)$
is in (\ref{forX}). The pole in $X(\beta)$ at $\beta=2\pi i/N$
means that particles in the representation $2\mu_2$ are the bound
state of two particles in the symmetric representation $2\mu_1$.
Because the factor $X(\beta)$ is the same as that of the $SU(N)$ 
Gross-Neveu model, the masses are the same:
$$m_a = m \sin(\pi a/N)$$ for the sigma model as well. However, the
multiplicites are different because the former are in representations
with highest weight $\mu_a$, while in the latter they  are in
representations with highest weight $2\mu_a$.

As discussed above, $\beta=-2\pi i /N$, the $S$ matrix
(\ref{vvsmat}) is entirely in the symmetric channel. Therefore, the
particles in the symmetric representation $2\mu_1$ can be viewed as
composites of those in the vector representation $\mu_1$. The same is
true for all the particles in the $SU(N)/SO(N)$ sigma model: those in
the representation $2\mu_a$ are composites of two particles in the
$\mu_a$ representation.  Because of this relation between
$S$ matrices, the transfer matrices are also related by
(\ref{tfused}) \cite{BRi,BRii}. 
This means that the resulting TBA systems
are closely related, and all the densities are labelled
in the same way. 
Explicitly, the Bethe equations for the $SU(N)/SO(N)$ sigma model
are obtained from those of the Gross-Neveu model
by two modifications. The kernel  $Y^{(N)}_{ab}$
coming from the $S$ matrix prefactor is replaced with
$Z^{(N)}_{ab}$, while the kernel $\sigma^{(\infty)}_{ab}$ is replaced
with $\tau^{(\infty)}_j$, defined by
\begin{equation}
 \tau^{(s)}_j (\beta) = \sigma^{(s)}_j (\beta+i\pi/N) +
\sigma^{(s)}_j (\beta-i\pi/N).
\label{tausigma}
\end{equation}
The sigma model version of  (\ref{betGN}) is
\begin{equation}
2\pi P_{a,0} (\beta) = {m_a}\cosh\beta + 
\sum_{b=1}^{N-1}Z^{(N)}_{ab}*\rho_{b,0}(\beta) - 
\sum_{j=1}^\infty
\tau^{(\infty)}_{j} * \widetilde{\rho}_{a,j}(\beta).
\label{betsig}
\end{equation}
while the  Bethe equations for the pseudoparticles are
\begin{equation}
2\pi {\rho}_{a,j}(\beta) = 
\tau_j^{(\infty)} * \rho_{a,0}(\beta) - \sum_{b=1}^{N-1}\sum_{l=1}^\infty 
A_{jl}^{(\infty)} * K_{ab}^{(N)} * \widetilde\rho_{b,l}(\beta)
\label{betsigj}
\end{equation}
Explicit expressions
for these kernels are given in the appendix.
Note how all these equations reduce to those of the sphere sigma model
by setting $N=2$.

The different kernels in the Bethe equations of course mean that the
TBA system is not quite the same as that of the Gross-Neveu
model. All the modifications involve the couplings of the functions of
$\rho_{a,0}(\beta)$ to the other $\rho_{b,j}$.  After using the
identities in the appendix, one finds that the net effect is to remove
couplings between $\epsilon_{a,0}$ to $\epsilon_{a,1}$ in the
Gross-Neveu TBA (\ref{tbaGN}), and replace them with a coupling
between $\epsilon_{a,0}$ to $\epsilon_{a,2}$. The $SU(N)/SO(N)$ TBA
equations are
\begin{eqnarray}
\nonumber
\epsilon_{a,j} (\beta) &=& 
T\sum_{b=1}^{N-1}
I^{(N)}_{ab}\int_{-\infty}^\infty \frac{d\beta'}{2\pi}
\frac{N}{2\cosh(N(\beta-\beta')/2)}
\ln\left(1+e^{\epsilon_{b,j}(\beta')}\right)\\
&&\qquad-T\sum_{l=0}^{\infty}
{\cal I}^{(\infty)}_{jl}\int_{-\infty}^\infty \frac{d\beta'}{2\pi}
\frac{N}{2\cosh(N(\beta-\beta')/2)}
\ln\left(1+e^{-\epsilon_{a,l}(\beta')}\right)
\label{tbasig}
\end{eqnarray}
The asymptotic conditions are the same as for the Gross-Neveu model.
In fact, the only difference is that the second incidence matrix
$I^{(\infty)}$ is replaced with ${\cal I}^{(\infty)}$.
These equations are displayed schematically in figure 4.

\begin{figure}
\centerline{\epsfxsize=2.5in\epsffile{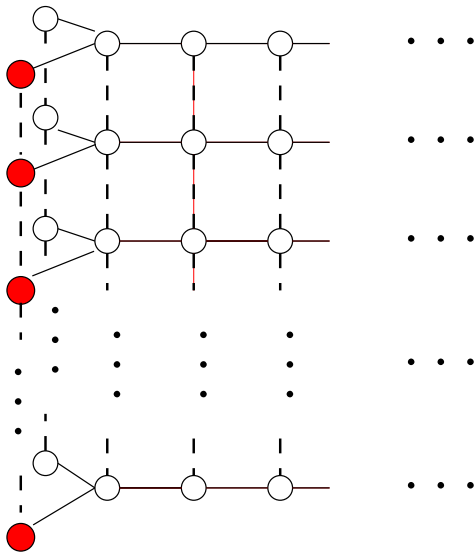}}
\bigskip\bigskip
\caption{The incidence diagram for the $SU(N)/SO(N)$ sigma model. There
are $N-1$ rows and an infinite number of columns.}
\end{figure}

Both cases can be conveniently summarized in the language of Dynkin
diagrams: the Gross-Neveu model in figure 3 is described by
($SU(N),SU(\infty)$), while the incidence diagram in figure 4 for the
$SU(N)/SO(N)$ sigma model is described by ($SU(N),SO(\infty)$). The
latter TBA system was previously discussed in \cite{tateo}, but
without the association with the sigma model.  As with all previous
cases, one can check that the UV and IR limits of the TBA equations
agree with known results, namely the central charge
$c_{UV}=(N+2)(N-1)/2$ and the particles being in the representations
$2\mu_a$.  This computation in particular checks that these are all
the particles in the spectrum, because additional (or fewer) particles
would change this central charge.

\subsection{$O(2P)$ Gross-Neveu models}

As with the models with $SU(N)$ symmetry, I will start with the
$O(2P)$ Gross-Neveu models \cite{GN,Witten,KT} (these are in
fact the models Gross and Neveu originally studied). Like the $SU(N)$
case, there are particles in all the fundamental representations with
highest weights $\mu_a$. This includes the spinor representations,
which physically correspond to kinks. The mass spectrum
is given by
$$m_a = m \sin(a\pi/(2P-2))\qquad\qquad m_{P-1}=m_P =
\frac{m}{2\sin(\pi/(2P-2))}$$ where the latter two correspond to the spinor
representations.  As opposed to the $SU(N)$ case, for $P\ge 4$ there
can be more than one representation with a given mass, as explained in
detail in \cite{KT}.  For any value of $a$ there are particles
in the representation $\mu_a$, but there may be additional ones as
well. For example, for $P=4$, there are particles in the vector
and spinor representations (all $3$ of them being $8$-dimensional)
of mass $m_1$,
particles in the antisymmetric representation ($28$-dimensional,
weight $\mu_2$) with mass $m_2 =\sqrt{3} m_1$, and a particle in the
singlet representation, with mass $m_2$. This apparently is related to
representation properties of the Yangian; it turns out that the
Yangian associated with $SO(8)$ has a $29$-dimensional representation,
but not a $28$-dimensional one. Under the $SO(8)$ subalgebra of the
Yangian, the $29$ decomposes into $28+1$.  In the TBA equations below,
the index $a$ indicates all particles of mass $m_a$, which presumably
corresponds to an irreducible representation of the Yangian \cite{Mackay}.

Luckily, the Bethe equations for $SO(2P)$-type systems were also found
in \cite{BRii,BRiii}. These were more or less conjectured based on
analogy with the $SU(N)$ case, but were proven up to some technical
assumptions in \cite{KNS}. Basically, they amount to doing
the computation by replacing the
$SU(N)$ incidence matrix $I^{(N)}$ with the $SO(2P)$ incidence matrix
${\cal I}^{(P)}$. The details for proving this are given in
the appendix. The TBA equations for the $O(2P)$ Gross-Neveu models are
\begin{eqnarray}
\nonumber
\epsilon_{a,j} (\beta) &=& 
T\sum_{b=1}^{P}
{\cal I}^{(P)}_{ab}\int_{-\infty}^\infty \frac{d\beta'}{2\pi}
\frac{P-1}{\cosh[(P-1)(\beta-\beta')]}
\ln\left(1+e^{\epsilon_{b,j}(\beta')}\right)\\
&&\qquad- T\sum_{l=0}^{\infty}
I^{(\infty)}_{jl}\int_{-\infty}^\infty \frac{d\beta'}{2\pi}
\frac{P-1}{\cosh[(P-1)(\beta-\beta')]}
\ln\left(1+e^{-\epsilon_{a,l}(\beta')}\right)
\label{tbaGNON}
\end{eqnarray}
These equations are displayed schematically in figure 5; the indices
$a$ and $b$ now run over the nodes of a $SO(2P)$ Dynkin diagram.
The correct central charge $c_{UV}=P$ is obtained in the UV limit.
This system
was also discussed in \cite{tateo}.

\begin{figure}
\centerline{\epsfxsize=2.5in\epsffile{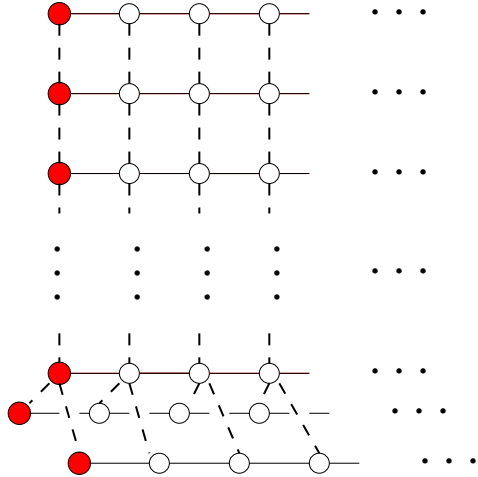}}
\bigskip\bigskip
\caption{The incidence diagram for the $O(2P)$ Gross-Neveu model. There
are $P$ rows and an infinite number of columns.}
\end{figure}

\subsection{$O(2P)/O(P)\times O(P)$ sigma models}

In \cite{metheta} the $O(2P)/O(P)\times O(P)$ sigma models were shown
to resemble the $SU(N)/SO(N)$ sigma models discussed above. This is
not terribly surprising, since the Lagrangian formulation of both is in
terms of symmetric matrix fields. I will show here
how their TBA systems are also similar.

In \cite{metheta} the exact spectrum and the $S$ matrix ${\cal
S}^{SS}$ for the $O(2P)/O(P)\times O(P)$ sigma models are found. Like
the $SU(N)/SO(N)$ case, there are particles are in all representations
with highest weight $2\mu_a$, where here $a=1\dots P$, although
because of some peculiarities of the $O(2P)$ $S$ matrices (and because
of Yangian representation properties), there must be particles in some
of the fundamental representations as well.  The sigma model mass
spectrum is the same as the $O(2P)$ Gross-Neveu model, although of
course the multiplicities differ.  The TBA system for the
$O(2P)/O(P)\times O(P)$ sigma models should not come as any surprise
at this point.  It follows from the $O(2P)$ Gross-Neveu model
calculation just as the $SU(N)/SO(N)$ calculation follows from that of
the $SU(N)$ Gross-Neveu model \cite{metheta}.  The TBA equations for
$O(2P)/O(P)\times O(P)$ sigma models are
\begin{eqnarray}
\nonumber
\epsilon_{a,j} (\beta) &=& 
T\sum_{b=1}^{P}
{\cal I}^{(P)}_{ab}\int_{-\infty}^\infty \frac{d\beta'}{2\pi}
\frac{P-1}{\cosh[(P-1)(\beta-\beta')]}
\ln\left(1+e^{\epsilon_{b,j}(\beta')}\right)\\
&&\qquad-T\sum_{l=0}^{\infty}
{\cal I}^{(\infty)}_{jl}\int_{-\infty}^\infty \frac{d\beta'}{2\pi}
\frac{P-1}{\cosh[(P-1)(\beta-\beta')]}
\ln\left(1+e^{-\epsilon_{a,l}(\beta')}\right)
\label{tbasigON}
\end{eqnarray}
The kernels and identities for this derivation are discussed in the
Appendix.

\section{Massless sigma models with $\theta=\pi$}

The results of the last section further confirmed the results of
\cite{metheta} for the $S$ matrices of the $SU(N)/SO(N)$ and
$O(2P)/O(P)\times O(P)$ sigma models when the instanton coupling
$\theta=0$. In this section, I find the TBA equations for these sigma
models when $\theta=\pi$, further confirming results of
\cite{metheta}.

The particles of the sigma models are massless when $\theta=\pi$. The
reason is that both sets of models have stable infrared fixed points,
the $SU(N)_1$ and $O(2P)_1$ WZW models, respectively. The $S$ matrices
for these flows were found in \cite{metheta}.  Since the particles are
massless, they are either left- or right-moving.  Rapidity variables
are still useful for parameterizing the energy and momentum of
massless particles: $E=p=m e^\beta$ for a right mover, and $E=-p = m
e^{-\beta}$ for a left mover. The parameter $m$ here is not the mass
of the particle, but rather is the scale (analogous to
$\Lambda_{QCD}$) which parameterizes the interactions. In
condensed-matter language, it is the crossover scale.  With these
definitions, the rapidity difference is still an invariant in a
collision. In a collision between a right mover and a left mover, the
invariant is $(E_1 + E_2)^2 - (p_1 + p_2)^2 = m^2e^{\beta_1 -
\beta_2}$. In ``collisions'' between two right movers, the invariant
is $E_1/E_2 = e^{\beta_1-\beta_2}$. I put collisions in quotes because
the $S$ matrix is properly interpreted here as a matching condition on
the wavefunction, as in (\ref{phaseshift}).  For more details on the
$S$ matrix approach to massless theories, see \cite{ZZtheta,FS}.

The spectrum and $S$ matrices of these sigma
models at $\theta=\pi$ are closely related
to that of the corresponding Gross-Neveu model.
For the
$SU(N)/SO(N)$ sigma model \cite{metheta},
\begin{eqnarray*}
S^{ab}_{LL}(\beta) &=& S^{ab}_{RR}(\beta) = S^{ab}_{GN} (\beta)\\
S^{ab}_{LR}(\beta) &=& S^{ab}_{GN} (\beta)/
X^{(N)}_{ab}(\beta)
\end{eqnarray*}
where $X^{(N)}_{ab}$ comes from fusing $X$ as defined in (\ref{forX}):
\begin{equation}
X^{(N)}_{ab}(\beta)\equiv \prod_{i=1}^{a}\prod_{j=1}^{b} 
 X\left(\beta+[i+j-1-(a+b)/2]/N\right).
\label{xab}
\end{equation}
For $N=2$, this reduces to the result of \cite{ZZtheta}.  The reason
for dividing out by $X^{(N)}_{ab}(\beta)$ in $S_{LR}$ is simple. Poles in
$S_{LR}$ in the region $0<Im(\beta)<\pi$ are forbidden
\cite{ZZtheta}, and all are contained in this factor.
For similar reasons, the $S$ matrices for the
$O(2P)/O(P)\times O(P)$ sigma model at $\theta=\pi$ are \cite{metheta}
\begin{eqnarray*}
{\cal S}^{ab}_{LL}(\beta) &=& {\cal S}^{ab}_{LL}(\beta) = {\cal
S}^{ab}_{GN} (\beta)\\ {\cal S}^{ab}_{LR}(\beta) &=& {\cal
S}^{ab}_{LR}(\beta) = {\cal S}^{ab}_{GN} (\beta)/ 
{\cal X}_{ab}^{(P)}(\beta)
\end{eqnarray*}
where 
\begin{equation}
{\cal X}_{ab}^{(P)}(\beta) = X^{(2P-2)}_{ab}(\beta)
X^{(2P-2)}_{ab}(i\pi-\beta)
\label{calX}
\end{equation}
and ${\cal S}^{ab}_{GN}$ is the $S$ matrix of the $O(2P)$ Gross-Neveu model.

The TBA systems follow from the results in the last section, given the
close relation with the Gross-Neveu models. The pseudoparticles
are identical, so the densities $\rho_{a,j}$ are
labeled by two indices as before.
However, in scattering, left
movers stay left moving, and right movers stay right moving. Thus 
instead of densities $\rho_{a,0}$, now there are
both $\rho_{a,L}$ and $\rho_{a,R}$.
For the $SU(N)$ case,
the first of the Bethe equations (\ref{betGN}) is replaced with
the two equations
\begin{eqnarray}
\nonumber
2\pi P_{a,R} (\beta) &=& {m_a}e^\beta +
\sum_{b=1}^{N-1}Y^{(N)}_{ab}*\rho_{b,R}(\beta) +  
\sum_{b=1}^{N-1}(Y^{(N)}_{ab}-\delta_{ab}\delta(\beta) 
+A^{(N)}_{ab})*\rho_{b,L}(\beta)  \\
&&\qquad -\sum_{j=1}^\infty
\sigma^{(\infty)}_{j} * \widetilde{\rho}_{a,j}(\beta)\\
\nonumber
2\pi P_{a,L} (\beta) &=& {m_a}e^{-\beta} +
\sum_{b=1}^{N-1}Y^{(N)}_{ab}*\rho_{b,L}(\beta) +  
\sum_{b=1}^{N-1}(Y^{(N)}_{ab}-\delta_{ab}\delta(\beta) +
A^{(N)}_{ab})*\rho_{b,R}(\beta)   \\
&&\qquad -\sum_{j=1}^\infty
\sigma^{(\infty)}_{j} * \widetilde{\rho}_{a,j}(\beta).
\label{betpi}
\end{eqnarray}
The Bethe equations for the pseudoparticles (\ref{betGNj}) become
\begin{equation}
2\pi {\rho}_{a,j}(\beta) = 
\sigma_j^{(\infty)} * (\rho_{a,L}(\beta)+\rho_{a,R}(\beta)) 
- \sum_{b=1}^{N-1}\sum_{l=1}^\infty 
A_{jl}^{(\infty)} * K_{ab}^{(N)} * \widetilde\rho_{b,l}(\beta)
\label{betpij}
\end{equation}
Using the identities in the appendix gives the TBA equations
\begin{eqnarray}
\nonumber
\epsilon_{a,j} (\beta) &=& 
\sum_{b=1}^{N-1}
I^{(N)}_{ab}\int_{-\infty}^\infty \frac{d\beta'}{2\pi}
\frac{N}{2\cosh(N(\beta-\beta')/2)}
\ln\left(1+e^{\epsilon_{b,j}(\beta')}\right)\\
&&\qquad-\sum_{l=L,R,1\dots\infty}
{\cal I}^{(\infty)}_{jl}\int_{-\infty}^\infty \frac{d\beta'}{2\pi}
\frac{N}{2\cosh(N(\beta-\beta')/2)}
\ln\left(1+e^{-\epsilon_{a,l}(\beta')}\right)
\label{tbasigpi}
\end{eqnarray}
where $j$ takes the values $L,R,1\dots\infty$. These equations 
for the $SU(N)/SO(N)$ sigma model at $\theta=\pi$ are
{\it identical} to those for the $SU(N)/SO(N)$ sigma model at
$\theta=0$ (\ref{tbasig}), once the labels are redefined (there $j$ is
takes the values $0,1\dots \infty$).  However, that does not mean
the solutions are the same. Because the $\theta=0$ theory is massive
and the $\theta=\pi$ theory is massless, the asymptotic conditions
are different. Namely, as $\beta\to \pm\infty$, for the massive theory:
\begin{eqnarray*}
\epsilon_{a0} (\beta\to \infty) &\longrightarrow& m_a \cosh(\beta)
\end{eqnarray*}
while for the massless theory as $\beta\to +\infty$
\begin{eqnarray*}
\epsilon_{aL} (\beta\to \infty) &\longrightarrow& m_a e^{\beta} \\
\epsilon_{aR} (\beta\to \infty) &\longrightarrow& constant
\end{eqnarray*}
and as $\beta\to -\infty$
\begin{eqnarray*}
\epsilon_{aL} (\beta\to -\infty) &\longrightarrow&  constant \\
\epsilon_{aR} (\beta\to -\infty) &\longrightarrow& m_a e^{-\beta}
\end{eqnarray*}
The free energy (\ref{free}) is modified in the massless case to
\begin{equation}
F^{(\pi)} (m,T)=
 - T \sum_a m_a \int_{-\infty}^{\infty} \frac{d\beta}{2\pi} 
\left[e^{\beta}
\ln\left( 1 + e^{-\epsilon_{aR}(\beta)/T}\right)
+ e^{-\beta}
\ln\left( 1 + e^{-\epsilon_{aL}(\beta)/T}\right)\right]
\end{equation}
The equations for the massless theory are pictorially depicted in figure 6.

\begin{figure}
\centerline{\epsfxsize=2.5in\epsffile{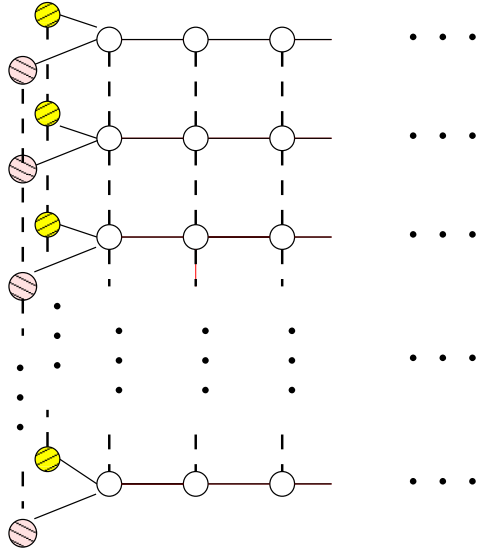}}
\bigskip\bigskip
\caption{The incidence diagram for the $SU(N)/SO(N)$
sigma model with $\theta=\pi$.
There are $N-1$ rows and an infinite number of columns.
The cross-hatched circles represent $\epsilon{aL}$ and $\epsilon_{aR}$.
}
\end{figure}

The different asymptotic conditions do not affect the free energy in
the ultraviolet limit $m/T \to 0$. Thus the free energy is the same in
massive and massless cases, corresponding to that of a conformal field
theory of central charge $c_{UV}=(N+2)(N-1)/2$. This of course is the
dimension of the manifold $SU(N)/SO(N)$.  In fact, because the TBA
systems are identical except for the asymptotic conditions, the entire
UV perturbation theory is identical in both cases.  This is as it must
be: instantons are a non-perturbative effect, and so the effect of the
instanton coupling $\beta$ cannot be seen in perturbation
theory. Unfortunately, it is not known how to compute the perturbative
expansion at non-zero temperature, except for the leading logarithmic
correction \cite{ZZtheta}. 
The perturbative expansion at zero temperature can be computed explicitly
by using a generalized Wiener-Hopf technique. This computation
was done for the case at hand in \cite{metheta}, and does indeed
give the same results at $\theta=0$ and $\pi$.

On the other hand, the physics for $\theta=\pi$ is radically different
from that at $\theta=0$ in the low-energy limit $m/T \to \infty$.
In the massive case the free energy in this limit is merely that of a
dilute gas of massive particles, as in (\ref{freeir}). However, the
particles are massless when $\theta=\pi$ because the system flows to a
non-trivial field theory in the low-energy limit. This flow is
immediately apparent from the $S$ matrix point of view, because the
two-particle Lorentz invariant for a left and a right mover is
$\propto m^2$, so the $S$ matrix goes to a $\beta$-independent
constant value as
$m\to\infty$. The
right-right and left-left matrices remain non-trivial, however, since
the Lorentz invariant here is independent of $m$. Thus in the
low-energy limit, the left and right sectors decouple from each other,
but remain non-trivial. This is the behavior of a conformal field
theory.  The free energy must obey a relation like that of the UV
limit, namely \cite{BCNA}
\begin{equation}
\lim_{m_a\to \infty} F = -\frac{\pi T^2}{6} c_{IR}
\label{freeirCFT}
\end{equation}
Here this gives $c_{IR}= N-1$. This is the central charge of
$SU(N)_1$, confirming the flow discussed in \cite{metheta}.  

In fact, since the left and right movers decouple in the IR limit, the
TBA system for the right movers in this limit is obtained merely by
removing the terms involving $\epsilon_{aL}$ from the equations.  The
resulting system is identical to that of the $SU(N)$ Gross-Neveu model
(\ref{tbaGN}); only the asymptotic condition changes from
$\epsilon_{a0}(\beta\to\infty) \rightarrow m_a\cosh\beta$ to
$\epsilon_{aR} (\beta\to\infty) \rightarrow m_a e^{\beta}$. The TBA
system for the left movers is the same, with the replacement $\beta\to
-\beta$. This close relation is a consequence of the fact discussed in
\cite{metheta}, that the effective field theory for the $SU(N)/SO(N)$
sigma model at $\theta=\pi$ in the low-energy limit is that of the
$SU(N)$ Gross-Neveu model at {\it negative} coupling.  The sign change
changes the sign of the beta function, meaning that while the
Gross-Neveu model is an asymptotically-free massive theory, the
critical point in the sigma model is stable. In another language, the
different signs correspond to marginally-relevant and
marginally-irrelevant perturbations respectively.

\bigskip

Not surprisingly, the $O(2P)/O(P)\times O(P)$ sigma model behaves in
the same fashion. The TBA system in (\ref{tbasigON}) applies to both
massive and massless cases. Only the asymptotic conditions differ, as
with the $SU(N)/SO(N)$ model. As a consequence, the same $c_{UV}=P^2$
is obtained for both $\theta=0$ and $\theta=\pi$.  In the massless
case, the flow is to a conformal field theory with $c_{IR}= P$, and
the equations in the IR limit are those of the $O(2P)$ Gross-Neveu
model. Thus indeed the flow is to the $O(2P)_1$ conformal field
theory, confirming the results of \cite{metheta}.

\section{Perturbed coset models}

In \cite{FZ,mecoset} it was shown how a $G/H$ sigma model is
related to a $G_k/H_l$ coset conformal field theory perturbed by a certain
operator. In this section, I review
this construction, and apply it to $G/H =
SU(N)/SO(N)$ and $O(2P)/O(P)\times O(P)$.
I find the exact free energy of the perturbed
coset models. this approach shows promise for understanding whether other
sigma models are integrable, as I will discuss in the conclusion.

\subsection{Perturbed coset models and sigma models}

A $G_k$ WZW model is a conformal field theory with an
infinite-dimensional symmetry algebra \cite{WZW,KZ}. 
This symmetry is an extension
of a ordinary Lie algebra symmetry $G$.
The symmetry currents
are denoted $J^A(z)$ and $\bar J^A(\bar z)$, where $A$ runs from $1\dots$
dim($G$). These currents satisfy the
operator product
\begin{equation}
J^A(z) J^B(w) = {k\over (z-w)^2} + {f^{ABC} J^C(w)\over z-w}+\dots
\label{OPE}
\end{equation}
where the $f^{ABC}$ are the structure constants of the ordinary Lie
algebra for $G$. The algebra (\ref{OPE}) is known as an affine Lie
algebra or a Kac-Moody algebra $G_k$.  The level $k$ is a positive
integer for a compact Lie group $G$.  The central charge (coefficient
of the conformal anomaly) of the $G_k$ WZW model is 
\begin{equation}
c=\frac{k\,\hbox{dim}\,G}{k+h}
\label{cwzw}
\end{equation}
where $h$ is called the dual Coxeter number.  It
can be defined by $f^{ACD}f^{BCD} = h \delta_{AB}/2$. For $G=SU(N)$,
$h=N$, while for $G=SO(2P)$, $h=2P-2$ (for $P>2$).  The primary
fields of the WZW model correspond to representations of $G_k$. It is
shown in \cite{KZ} that they have scaling dimensions 
\begin{equation}
x_j=\frac{2C_j}{(k+h)}
\label{xwzw}
\end{equation}
where $C_j$ is the quadratic Casimir defined by $T^A T^A= C_j {I}$, 
with the $T^A$ the generators of the Lie algebra of $G$ in the
$j$th representation and ${I}$ the identity matrix. All the other
scaling fields arise from the operator product of the $J^A(z)$ with
the primary fields; it follows from (\ref{OPE}) that $J$ has dimension
$1$ and therefore all fields have dimensions $x_j$ plus an integer.

A coset conformal field theory $G_k/H_l$ is formed from a $G_k$ WZW
theory and a subalgebra $H_l$.  The energy-momentum tensor is
constructed from the generators of $G_k$ not in $H_l$
\cite{GKO}.  The central charge of this new conformal field theory is
$c(G_k) - c(H_l)$. The level $l$ of the subalgebra $H$ is determined
is given by $l=kr$, where
$r$ is a group-theory factor called the index of the embedding of $G$
into $H_l$. For the embedding of $SO(N)$ into $SU(N)$, $r=2$ ($r=4$ for
$N$=3), while for the embedding of $O(N)\times O(N)$ into $O(2N)$,
$r=1$ ($r=2$ for $N=3$).

The fields of the $G_k/H_l$ conformal field theory are
constructed by decomposing a field $\phi_G$ in $G_k$ into
representations of the $H_l$ subalgebra. Because the energy-momentum
tensor obeys the orthogonal decomposition $T_G=T_H + T_{G/H}$, the
decomposition of $\phi_G$ must be of the form
\begin{equation}
\phi_G = \oplus_a  \phi^a_{G/H} \otimes \phi^a_H.
\label{decomp}
\end{equation}
The coefficients $\phi^a_{G/H}$ of this decomposition are the fields
of the coset model $G_k/H_l$.

These coset conformal field theories {\it a priori} have nothing to do
with $G/H$ sigma models. The former are massless, and do not have a
global symmetry $G$, while the latter are gapped with a $G$ global
symmetry. Thus for the two to correspond, the coset model must be
perturbed by some operator. Moreover, the coset model has a $G$ global
symmetry when $k\to\infty$. These and other considerations led to a
conjecture made in \cite{mecoset}. This conjecture is that the sigma
model for $G/H$ is equivalent to the $k\to\infty$ limit of the coset
conformal field theory perturbed by a certain operator.  The operator
is obtained by using (\ref{decomp}) to decompose the currents $J^{A}$
into fields in $G_k/H_l$.  For the cases of interest here, $G/H$ is a
symmetric space, meaning that there is no normal subgroup of $G$
containing $H$ other than $G$ itself.  A consequence of $G/H$ being a
symmetric space is that the generators of $G$ not in $H$ form a real
irreducible representation of $H$ \cite{ZJ}.  Thus when a field
$J^A(z)$ is decomposed into representations of $H$ in (\ref{decomp})
there is only one term on the right hand-side. The resulting field in
$G_k/H_l$ is denoted by ${\cal J}^A$. The fields ${\cal J}^A$ form a
real irreducible representation of $H$, of dimension
$c_{UV}={\hbox{dim}\,G}-{\hbox{dim}\, H}$.  The operator ${\cal O}_\sigma$ is
defined as
\begin{equation}
{\cal O}_{\sigma} \equiv \sum_{A =1}^{c_{UV}}
{\cal J}^{A}(z) {\cal J}^{A}(\overline z).
\label{pert}
\end{equation}
The conjecture of \cite{mecoset} can now
be stated precisely: the $G/H$ sigma model is equivalent
to the $G_k/H_l$ coset conformal field theory perturbed by the operator
${\cal O}_\sigma$ in the limit $k\to\infty$.

The conjecture passes a few simple checks.  The ultraviolet limit is
obtained by removing the perturbation of the coset model.  From (\ref{cwzw})
it follows that the central charge of the $G_k/H_l$ theory as $k\to
\infty$ is indeed $c_{UV}=\hbox{dim}\,G - \hbox{dim}\,H$ as in the
sigma model. Moreover, when one decomposes $J^A$ into representations
of $H$ for $A$ in $G$ but not $H$, the resulting field $\phi_H^A$ has
dimension going to zero as $k\to\infty$, because the quadratic Casimir
in (\ref{xwzw}) is independent of $k$. Thus the field ${\cal J}^A$ has
dimension $1$ in this limit, so the perturbation ${\cal O}_\sigma$ is
of dimension $2$ and so is naively marginal. It is not exactly
marginal -- this is the phenomenon of dimensional transmutation and
asymptotic freedom.  Therefore the coset and its perturbation have the
general properties of a sigma model.  Further support for this
conjecture is discussed in \cite{mecoset}.  For example, it has been
shown to be true for the principal chiral models \cite{ABL}, and in
the sphere sigma model \cite{FZ}.
The results in this section give strong further evidence in
support.

The models of interest in this paper are the
$SU(N)_k/SO(N)_{2k}$ and the $O(2P)_k/O(P)_k\times O(P)_k$ 
conformal field theories perturbed by ${\cal O}_\sigma$.
The former theories have
$$c_{UV}(k,N) = \frac{k(k-1)(N+2)(N-1)}{(N+k)(N-2+2k)}$$
while the latter have
$$c_{UV}(k,P) = \frac{k(k-1)P^2}{(P-2+k)(2P-2+k)}$$
To find the dimensions of the perturbing operators requires a little group
theory. Fields in the adjoint representation of $SU(N)$ decompose under
the $SO(N)$ subgroup as
$$(N^2-1) \to \left(\frac{N(N-1)}{2}\right) + \left(
\frac{N(N+1)}{2} -1\right)$$
The representation of dimension $N(N-1)/2$ consists of
the generators $J^A$
with $A$ in the $SO(N)$ subgroup as well. Thus the operators ${\cal J}^A$
are in the symmetric representation of $SO(N)_{2k}$, of 
dimension $N(N+1)/2 - 1$. The quadratic Casimir of this representation
is $C_{sym} = N$. 
Since the dimension of $J^A$ is always $1$, the dimension
of ${\cal O}_\sigma$ in the $SU(N)_k/SO(N)_{2k}$ 
conformal field theory for $N>2$ is
$$x_{\sigma} = 2- \frac{2N}{N-2+2k}=4\frac{k-1}{N-2+2k}$$
Similarly, the adjoint representation of $O(2P)$ decomposes
into
$$(P(2P-1))\to \left(\frac{P(P-1)}{2},1\right) +
 \left(1,\frac{P(P-1)}{2}\right) + (P,P)$$
under the $O(P)\times O(P)$ subgroup.
Thus the operator ${\cal O}_\sigma$ here is in the $(P,P)$ representation
of $O(P)_k\times O(P)_k$. The quadratic Casimir of the vector representation
of $O(P)$ is $(P-1)/2 $, so
\begin{equation}
x_{\sigma} = 2 - \frac{P-1}{P-2+k}=2\frac{k-1}{P-2+k}
\label{lam}
\end{equation}
As far as I known, these perturbed conformal field theories have
never been studied in the literature.

The role the instanton coupling $\theta$ takes in the conjecture of
\cite{FZ,mecoset} is quite interesting. The action of the perturbed
conformal field theories can be denoted schematically as $$S= S_{CFT}
+ \lambda \int d^2 z\, {\cal O}(z,\bar z).$$ It follows from simple
scaling considerations that mass scale $m$ in the theory is related to
$\lambda$ by $m\propto |\lambda|^{1/(2-x)}$. If the theory has a ${\bf
Z}_2$ symmetry under which ${\cal O}\to - {\cal O}$ then the theories
with positive and negative $\lambda$ are identical. In general, they
are not.  A well known example is the $SU(2)_k\times
SU(2)_1/SU(2)_{k+1}$ ``minimal'' models of conformal field theory
perturbed by ${\cal O}_\sigma$ (usually called $\phi_{1,3}$ in this
context). With one sign of $\lambda$, the model is massive.  With the
other sign, the model flows to the minimal model with $k-1$ \cite{LC},
so the excitations are massless. 
In the $SU(N)_k/SO(N)_{2k}$ and
$O(2P)_k/O(P)_k\times O(P)_k$ cases for $k>2$, the two signs of $\lambda$ give
different theories as well, one massive and the other massless. In the
$k\to\infty$ limit these differing theories correspond to $\theta=0$ and
$\theta=\pi$ respectively. This was argued in \cite{FZ} for
$SU(2)_k/O(2)$.
Strikingly, one can also see from
the perturbed conformal field theories here that that the different sign
affects perturbation theory only at the order $\lambda^{k}$. Thus
as $k\to\infty$, the different sign does not affect perturbation
theory. Its only effects are non-perturbative, just as they must be
if the change $\lambda\to -\lambda$
is to describe the effects of a $\theta$ term.

\subsection{The particle spectrum}

Here I discuss the particle spectrum of the perturbed conformal field
theories just defined.

The results for the simplest cases $k=2$ are already well known
(when $k=1$, the models are trivial).
The $SU(N)_2/O(N)_4$ conformal field
theories are known as the $Z_N$ parafermion theories; the equivalence
to the better known coset description $SU(N)_1\times SU(N)_1/SU(N)_2$
was shown in \cite{Alt}. The perturbation
${\cal O}_\sigma$ of dimension $2/(N+2)$
is called the thermal operator here. This
is an integrable field theory, with $S$ matrices derived in \cite{KS}.
The spectrum consists
of $N-1$ particles, with mass \cite{KM}
$$m_a = m \sin\left(\frac{a\pi}{N}\right)$$ This is the same mass
spectrum as in the $SU(N)$ Gross-Neveu model and the $SU(N)/SO(N)$
sigma models discussed above. The degeneracies are different: there is
only one particle of each mass in the parafermion model, while in the
other cases, there are multiplets of particles in $SU(N)$
representations with highest weights $\mu_a$ and $2\mu_a$
respectively. The parafermion theory has a ${\bf Z}_N$ symmetry, but
no $SU(N)$ symmetry. 

Likewise, the $O(2P)_2/O(P)_2\times O(P)_2$ conformal field theories
are the ${\cal D}_{2P}$ parafermion theories. 
Their symmetry group is not $O(2P)$, but
instead the dihedral group ${\cal D}_{2P}$.
The equivalence to the usual
formulation $O(2P)_1 \times O(2P)_1/O(2P)_2$ formulation of these
parafermion theories can be shown using the techniques of
\cite{Alt}. These theories have $c=1$ for any $P$. The
perturbation is of dimension $1/P$, and so the massive theory
corresponds to the sine-Gordon model at $\beta^2 = 8\pi/P$. This is of
course integrable \cite{ZZ}, and in fact corresponds to the
``reflectionless'' points of sine-Gordon, where the scattering is
diagonal. The spectrum consists of $P$ particles, of masses
$$m_a = m \sin(a\pi/(2P-2))\qquad\qquad m_{P-1}=m_P =
\frac{m}{2\sin(\pi/(2P-2))}$$ The particles of masses $P$ and $P-1$
are the kink and antikink of the sine-Gordon model. This mass spectrum
is the same as that in the $O(2P)$ Gross-Neveu model, and the
$O(2P)/O(P)\times O(P)$ sigma model, but with
multiplicity 1 here.

The fact that the mass spectrum of the $k=2$ perturbed coset models
are the same as the corresponding sigma models is already a strong
piece of evidence in support of the conjecture of \cite{mecoset}.
The issue now is to find the spectrum and $S$ matrices for general $k$.
For $N=2$ and $P=2$, the answers are given in \cite{Fateev}, but otherwise
these models have not been discussed in the literature.
I will solve this problem for all $k$.

To understand the particle spectrum in an integrable model, it is
crucial to understand the symmetries of the model. For the sigma
models, this symmetry algebra is an ordinary Lie algebra $G$.  I
conjecture that the perturbed coset models are invariant under a
one-parameter deformation of $G$ called the {\it quantum-group algebra}
$U_q(G)$.  The particles in the perturbed $G_k/H_l$ models form
finite-dimensional representations of $U_q(G)$, with the parameter
$q=e^{i\pi/(k+h)}$, where $h$ is still the dual Coxeter number.  As
$k\to\infty$, $q\to 1$ and the algebra reverts to the usual $G$ Lie
algebra.  All known integrable perturbations of coset conformal field
theories are proven or believed to be invariant under some such
quantum-group algebra. For example, for models where $G=H\times H$,
this was discussed in detail in \cite{ABL}. For other models, this was
discussed in \cite{Vaysburd}. For the cases of interest here, the
particles in the $SU(2)_k/O(2)$ were shown to form a representation of
$U_q(SU(2))$ in \cite{Fateev}.

To give a concrete example, $U_q(SU(2))$ is the algebra
\begin{equation}
[S_z,S_\pm]=\pm 2S_\pm,\quad
[S_+,S_-]={q^{2S_z}-q^{-2S_z}\over q-q^{-1}}
\label{su2q}
\end{equation}
When the parameter $q=1$, this reverts to the usual $SU(2)$ Lie
algebra.  A nice physical realization of this algebra is
discussed in \cite{PS}, where it is shown how when the Heisenberg spin
chain is deformed into the XXZ spin chain, the $SU(2)$ symmetry is
deformed into $U_q(SU(2))$. 
The properties of the representations of $U_q(G)$
can be quite different from those of $G$ when $q$
is a root of unity other than $1$ or
$-1$. For
example, the right-hand-side of the last equation in (\ref{su2q})
vanishes on states with $2S_z= p$ when $q^p=1$. This means that
representations with maximum value of $2S_z$ greater than $p$ are
reducible. In other words, the only irreducible representations have
$|2S_z|<p$, as opposed to ordinary $SU(2)$, where there are
irreducible representations with any integer value of $2S_z$.

Particles in a representation of a quantum-group algebra are most
conveniently treated as {\it restricted kinks} \cite{ZRSOS}. Consider a field
$\phi$, with a potential $V(\phi)$ tuned so that there are degenerate
minima, which I will sometimes call vacua. Then kinks are field
configurations with $\phi(x=-\infty)$ one minimum of the potential,
$\phi(x=\infty)$ another. The kinks in the perturbed coset models form
what are called ``restricted solid-on-solid'', or RSOS,
representations of the quantum-group algebra.  The name comes from the
statistical mechanical lattice models in which these representations
first arose \cite{ABF}.

For $U_q(SU(2))$, these restricted kinks are easy to describe. They
interpolate between the minimum of a potential which has $k+1$ minima
in a row.  For example, the potential $V(\phi)=\phi^2 (\phi^2-1)^2$
has three minima at $\phi=0,\pm 1$; the potential
$V(\phi)=(\phi^2-1)^2(\phi^2-9)^2$ has four vacua in a row.  Kinks in
these sorts of potentials provide representations of the quantum-group
algebra $U_q(SU(2))$ with $q$ a root of unity. The two-dimensional
representations are kinks which interpolate between adjacent vacua.
Such representations behave just like ordinary $SU(2)$ spin-$1/2$
representations.  For example, for $k=2$, there are three minima
labeled $0,\pm 1$, and the generators $S_\pm$ exchange the states
$\phi(-\infty)=0$ and $\phi(+\infty)=\pm 1$.  To construct the larger
representations, one can take the tensor product of smaller
representations. The rules are just like that of ordinary $SU(2)$: for
example, the tensor product of two spin-$1/2$ representations
decomposes into the sum of a spin-$1$ and a spin-$0$
representation. The one catch is that for $q$ a root of unity, the
larger representations are reducible.  For $k=2$, spin $0,1/2$ and $1$
are all the irreducible representations. This is clearly apparent from
the kink picture, because for $k=2$ there are only three vacua: an
irreducible spin-$3/2$ representation requires four vacua.  Moreover,
even the allowed kinks are restricted. Restricted means that
multi-kink configurations must obey the rules implied by the
potential. The number of ${\cal N}$-kink states is much less than the
number of one-kink states to the ${\cal N}^{\hbox{th}}$ power. For
$k=2$, in fact, there is only one way to construct a multi-particle
state from spin-$1$ particles: the vacua must alternate between $+1$
and $-1$. The
restriction is so strong that the kink structure gives no new degrees
of freedom, so it can be viewed as a normal particle.

Perturbed coset models with restricted kinks are already widely
known. The $SU(2)_{k}\times SU(2)_1/SU(2)_{k+1}$ minimal models
perturbed by ${\cal O}_\sigma$ are integrable.  The particles are
spin-$1/2$ $U_q(SU(2))$ kinks, where $q=e^{i\pi/(k+2)}$
\cite{ZRSOS,ABL}. There are thus $k+1$ vacua here, with the kinks
interpolating between adjacent vacua.  The $k=1$ case corresponds to
the thermal perturbation of the Ising model (free Majorana fermions).
Since there are only two wells when $k=1$, all the kink can do is go
back and forth, and one can forget it is a kink.  For the case
$SU(2)_k/O(2)$, the particles are spin-$1$ $U_q(SU(2))$ kinks
\cite{Fateev}. The $k=2$ case here also corresponds to the thermal
perturbation of the Ising model. In this description, there are three
vacua, but the kinks are of spin $1$, so again all they can do is go
back and forth: there is only one state for a given number of
particles.

For general algebras $U_q(G)$, the restricted-kink structure is more
complicated. The potential is defined so that the minima correspond to
the highest-weight states of the quantum-group algebra allowed at that
value of $q$. For simply-laced algebras, the allowed weights $\sum_a
c_a\mu_a$ must satisfy $\sum_a c_a \le k$.  In this language, for
$U_q(SU(2))$ with $q^4=1$ ($k=2$), the three minima correspond to
highest weights $0,\mu_1,2\mu_1$, where $\mu_1$ is the sole
fundamental weight of $SU(2)$.  The kinks form representations of the
algebra, so each kink is also labelled by a weight.  The rule is then
that there can be a kink of representation $r_a$ interpolating from
the vacuum $\gamma$ to the vacuum $\delta$ if the corresponding representations
obey the tensor product
$$r_{a}\otimes r_{\gamma} = r_{\delta} \oplus \dots$$ 
I said ``can be'' because it depends on the specifics of a given
theory if such a kink actually does appear in the spectrum.  There are
a number of subtleties with this picture for general groups and
representations, but it is not necessary to understand them for this
work.

Given a particle spectrum consisting of restricted kinks, the $S$
matrix can be found using the Boltzmann weights of the corresponding
lattice statistical-mechanical model, which is usually known as the
$R$ matrix.  For models with particles in the fundamental
representations, this was discussed in \cite{ZRSOS,ABL,dVF}.  I
emphasize that by corresponding lattice model, I do {\it not} mean a
lattice model whose continuum limit is described by a field theory
with this $S$ matrix.  I mean that there is some integrable lattice
model whose Boltzmann weights are proportional to the $S$ matrix.  In
the corresponding lattice models, the variables which placed on sites
of the lattice play the role of the vacua, while the kinks correspond
to the states on the links.  The rapidity difference in the $S$ matrix
corresponds to the spectral parameter in the lattice model. The
scattering of kinks in representation $a$ from one in representation
$b$ is given by the matrix $S^{ab} \propto R^{ab}$ (as before, $a$ and
$b$ are not the matrix indices, but rather label the different
matrices).  The prefactor is not of interest to the lattice model,
since it merely multiplies the partition function by an overall
factor.  It is of course of great importance to the $S$ matrix theory.
The $R$ matrices for the $RSOS$ models are trigonometric solutions of
the Yang-Baxter equation. They can be written in the form
(\ref{project}), where the $f^{ab}_c$ are trigonometric functions (as
opposed to the rational functions appearing in the sigma models).
They are given explicitly for the fundamental representations of all
the quantum-group algebras $U_q(G)$ in \cite{KyotoR}, generalizing the
$SU(2)$ results of \cite{ABF}. The fusion procedure also can be used
to construct the $R$ matrices for kinks in the representations
$2\mu_a$ \cite{Kyotofus}.

The spectrum of the perturbed coset models is easy to obtain, given
the sigma model result. The kinks must be in the same representation
of $U_q(G)$ as the particles are of $G$.  For example, for the case
$SU(2)_k/U(1)$, the kinks are in the spin-$1$ representation of
$U_q(SU(2))$, while the particles in the sigma model are in the
spin-$1$ representation of $SU(2)$.  When $k$ is finite, the vacua are
restricted, but the restriction is removed as $k\to\infty$: the
particles in the sigma model no longer need be viewed as
kinks. Similarly, for the massive perturbed $SU(N)_k/SO(N)_{2k}$ and
$O(2P)_k/O(P)_k\times O(P)_k$ models, the kinks are in all
representations $2\mu_a$ for $a=1\dots N-1$ and $a=1\dots P$
respectively.  The vacua are all weights $\sum_a c_a \mu_a$ with
$\sum_a c_a \le k$.

Note also that the $H$ Gross-Neveu model is obtained by taking
$k\to\infty$ in the perturbed coset models $H_{k}\times H_1/H_{k+1}$
\cite{ABL,dVF,Hollo}.

\subsection{The free energy of the perturbed coset models}

The derivations of the TBA equations for the perturbed coset models
requires diagonalizing the transfer matrices formed from the kink
$S$ matrices. 
The computation is
very similar for those of the sigma models, because the
analysis of \cite{BRii,BRiii} applies to the RSOS models.

It is simplest to first discuss the case $k=2$, where
the perturbed coset models reduce to the well-studied
parafermion theories. As explained above, the kink structure is trivial:
there is only one particle for each representation $a=1\dots N-1$ or
$a=1\dots P$. The scattering here is
diagonal but non-trivial. The $S$ matrix element for scattering a
particle of type $a$ from one of type $b$ for $SU(N)$ parafermions is
$$S^{ab}(\beta) = X^{(N)}_{ab}(\beta)$$ where $X^{(N)}_{ab}(\beta)$ is
defined in (\ref{xab}).  For the $O(2P)$ parafermions, the $S$ matrix
elements are ${\cal X}^{(P)}_{ab}(\beta)$, as defined in (\ref{calX}).
The TBA equations instantly follow from using these $S$ matrices to
give the kernels in (\ref{tba}). There are no pseudoparticles because
the scattering is diagonal, so the only functions which appear can be
labelled $\epsilon_{a,0}$. Note the distinction with the sphere sigma
model, where the only functions which appear are $\epsilon_{1,j}$ in
the present notation.  For the $SU(N)$ parafermions, the TBA equations
are \cite{KM}
$$\epsilon_{a,0}(\beta) = m_a \cosh\beta 
- T\ln\left(1+ e^{-\epsilon_{a,0}(\beta)/T}\right)
+T \sum_{b=1}^{N-1} \int_{-\infty}^\infty
\frac{d\beta'}{2\pi} A_{ab}^{(N)}(\beta-\beta') 
\ln\left(1+e^{-\epsilon_{b,0}(\beta')/T}\right)
$$
where $A_{ab}^{(N)}$ is the same kernel which appeared in the Bethe
equations above, and is given explicitly in the appendix.  This can be
simplified greatly by using the fact that $A$ and $K$ are inverses,
giving
\begin{equation}
\epsilon_{a,0}(\beta) = T \sum_{b=1}^{N-1} I_{ab}^{(N)} \int_{-\infty}^\infty
\frac{d\beta'}{2\pi}\frac{N/2}{\cosh[N(\beta-\beta')/2]}
\ln\left(1+e^{-\epsilon_{b,0}(\beta')/T}\right)
\end{equation}
 where the asymptotic
condition $\epsilon_{a,0}\to m_a\cosh\beta$ as $\beta\to\infty$ is
implied.  The incidence matrix couples only ``adjacent'' functions;
it is displayed by restricting the diagram in figure 3 or 4 to have
only one column.
For the $O(2P)$ parafermions, the kernel
$A_{ab}^{(N)}$ is replaced by ${\cal A}_{ab}^{(P)}$ \cite{KM}. This results
in the TBA equations
\begin{equation}
\epsilon_{a,0}(\beta) = 
T \sum_{b=1}^{P} {\cal I}_{ab}^{(P)} \int_{-\infty}^\infty
\frac{d\beta'}{2\pi}\frac{P-1}{\cosh[(P-1)(\beta-\beta')]}
\ln\left(1+e^{-\epsilon_{b,0}(\beta')/T}\right)
\end{equation}
Thus the TBA equations for the
$k=2$ cases amount to those of the corresponding
sigma models with all the pseudoparticles
removed.

The TBA equations for general $k$ are also found by truncating the
equations for the corresponding sigma model. The reason is simple to
describe schematically.  Each irreducible representation of the
quantum-group algebra is associated with some transfer matrix.
Relations like the fusion relation (\ref{tfused}) relate the different
transfer matrices. The fact that there are only a finite number of
irreducible representations of the quantum-group algebra means that
the fusion relations relating all these transfer matrices truncate
\cite{BRi,BRii,BRiii}.  In the Bethe ansatz equations, this means that
there are only a finite number of pseudoparticles. In the TBA
equation, the index $j$ in the functions $\epsilon_{a,j}$ now runs
only from $0\dots k-1$ in the massive case.

This derivation of the 
Bethe equations is covered in detail
in \cite{BRii,BRiii,Hollo}. For the $SU(N)$ case,
for example, the Bethe equations for the pseudoparticles
are very similar to (\ref{betGNj}), but are modified to
\begin{equation}
2\pi {\rho}_{a,j}(\beta) = 
\sigma_j^{(k)} * \rho_{a,0}(\beta) - \sum_{b=1}^{N-1}\sum_{l=1}^{k-1}
A_{jl}^{(k)} * K_{ab}^{(N)} * \widetilde\rho_{b,l}(\beta)
\label{betcosetj}
\end{equation}
The equations for the $O(2P)$ case are modified in a similar fashion.
The $S$ matrix prefactor is modified as well; the kernel for
the $SU(N)_k/SO(N)_{2k}$ case is given in the appendix.

The result of these modifications is that the
TBA equations are truncated. Like the TBA equations
(\ref{tbaGN},\ref{tbasig},\ref{tbaGNON},\ref{tbasigON})
they are of the form
\begin{eqnarray}
\nonumber
\epsilon_{a,j} (\beta) &=& 
T\sum_{b=1}^{\hbox{rank}\, G}
Q_{ab}\int_{-\infty}^\infty \frac{d\beta'}{2\pi}
\frac{h}{2\cosh[h(\beta-\beta')/2]}
\ln\left(1+e^{\epsilon_{b,j}(\beta')}\right)\\
&&\qquad- T\sum_{l=0}^{k-1}
R_{jl}\int_{-\infty}^\infty \frac{d\beta'}{2\pi}
\frac{h}{2\cosh[h(\beta-\beta')/2]}
\ln\left(1+e^{-\epsilon_{a,l}(\beta')}\right)
\label{tbaall}
\end{eqnarray}
where $h$ is the dual Coxeter number for $G$, which is $N$ for $SU(N)$,
and $2P-2$ for $O(2P)$. The rank of $SU(N)$ is $N-1$, and the rank
of $O(2P)$ is $P$.
The matrices $Q$ and $R$ are all incidence matrices. For the various models
considered here, the results are given in the following table.
\begin{center}
\begin{tabular}{|c||c|c||c|}\hline
coset model perturbed by ${\cal O}_\sigma$
&$Q$&$R$&behavior when $k\to\infty$\\ \hline\hline
$SU(N)_k\times SU(N)_1/SU(N)_{k+1}$
&$I^{(N)}$&$I^{(k)}$&$SU(N)$ Gross-Neveu model\\
\hline
$SU(N)_k/SO(N)_{2k}$
&${I}^{(N)}$&${\cal I}^{(k)}$&$SU(N)/SO(N)$ sigma model\\
\hline
$O(2P)_k\times O(2P)_1/ O(2P)_{k+1}$
&${\cal I}^{(P)}$&${I}^{(k)}$&$O(2P)$ Gross-Neveu model\\
\hline
$O(2P)_k/O(P)_k\times O(P)_k$
&${\cal I}^{(P)}$&${\cal I}^{(k)}$&$O(2P)/O(P)\times O(P)$ sigma model\\
\hline
\end{tabular}\\ \smallskip \end{center}
In all cases, the usual asymptotic conditions apply.
All the TBA equations in this paper are contained in this table.
One can check that the central charges resulting
from taking the UV limit of the TBA equations are indeed those of the
corresponding conformal field theories for any value of $k$. This
is an enormous check on all the results of this paper.

\subsection{Flows between coset models}

I showed for the sigma models that the TBA equations for
$\theta=0$ and $\theta=\pi$ are identical, with the only difference
being in the asymptotic conditions.  
The same behavior should happen
for the two signs of $\lambda$ in the perturbed coset models (\ref{lam}). 
The TBA results for the perturbed coset models make it possible to 
understand the flow when the perturbation is massless.
The TBA equations (\ref{tbaall}) and the table still hold,
except that the asymptotic conditions given in section 4 apply here.
The sum over $l$ now runs from $L,R,1\dots k-2$.  
The IR fixed point can be read off from the equations, as described
above for the sigma models. Removing say the left moving particles
from the $SU(N)_k/SO(N)_{2k}$ perturbation gives the diagram for
the $SU(N)_{k-1}\times SU(N)_1/SU(N)_{k}$ models. Thus the flow is
between the conformal field theories
$$\frac{SU(N)_k}{SO(N)_{2k}} \longrightarrow \
\frac{SU(N)_{k-1}\times SU(N)_1}{SU(N)_{k}}$$
Likewise there is a flow
$$\frac{O(2P)_k}{O(P)_k \times O(P)_k} \longrightarrow
\frac{O(2P)_{k-1} \times O(2P)_1}{O(2P)_k}$$ As far as I know, these
flows were previously unknown. 
When $k=2$, there is no flow: the two cosets are already equivalent.
By using the equivalences between
different coset models derived in \cite{Alt}, one can described these
flows in different ways, if desired. For example, the latter also amounts
to a flow
$$\frac{O(k)_P\times O(k)_P}{O(k)_{2P}} \longrightarrow
\frac{O(2P)_{k-1} \times O(2P)_1}{O(2P)_k}.$$

Going backwards, one can read off the spectrum and particles
for these massless perturbations.
The kinks must be massless,
and in all representations $\mu_a$, and are either left or
right-moving. The vacua correspond to all weights
$\sum_a c_a \mu_a$ with $\sum_a c_a \le k-1$.
This shift of $k\to k-1$ indicates the quantum-group parameter $q$ is
different for the massless and massive perturbations, but I do not
know the reason for this. In the coset models $H_k\times H_1/H_{k+1}$
there are two quantum-group symmetries for both perturbations
\cite{ABL}; presumably the same thing happens here.

\section{Conclusion}

In this paper I have described how to compute the exact free energy
in integrable two-dimensional sigma
models. This definitively establishes that when $\theta=\pi$, there
are non-trivial fixed points for two sets of sigma models. It also
yields the exact free energy and susceptibility when $\theta=0$ and
when $\theta=\pi$.

The big open question is if other sigma models are integrable. The
grail in particle physics is probably the
$CP^{N-1}=SU(N)/SU(N-1)\times U(1)$ models. They have been widely
studied because they allow instantons and are tractable in large $N$. 
(The models studied above have a parameter $N$ and have
instantons, but they are difficult to treat in large $N$. The reason
is that they are matrix fields: the number of fields at large $N$
grows as $N^2$, not as $N$.) In particular, the $CP^N$ models allowed
Witten to conclude that instantons were not important in real-world
QCD \cite{WittCP}. It would be very interesting to prove Witten's
results directly, instead of relying on large $N$.

Virtually all the symmetric-space sigma models have arisen in various
condensed-matter applications \cite{Zirn}, but the grail here is the
$U(2N)/U(N)\times U(N)$ ``Grassmanian'' model. The reason is that in
the replica limit $N\to 0$, this is believed to describe the
transition between quantum Hall plateaus \cite{Pruisken}. This
transition is experimentally realized, and good numerical and
experimental measurements have been made of critical exponents. These
critical exponents should arise in some conformal field theory, but it
is still not known which one. Solving the sigma model as a function of
$N$ would presumably solve this problem.

So why are sigma models integrable? In some sigma models (see e.g.\
\cite{Luscher, Abdalla}), one can find non-local conserved currents.
Although the existence of non-local currents does not prove
integrability, it is a good indicator.  Often these non-local currents
are often associated with quantum-group or Yangian symmetry algebras.
In the $O(N)/O(N-1)$ models, one can prove the non-local currents
of \cite{Luscher} are the generators of an infinite-dimensional
symmetry algebra called the Yangian \cite{Denis}.  This proves the
integrability of these sigma models.  Unfortunately, this result has
not yet been extended to other sigma models.

So are other sigma models integrable?  An old result (see e.g.\
\cite{Abdalla}) suggested that the only integrable symmetric-space
$G/H$ sigma models are those where $H$ is a simple Lie group. The
reason is that they found that the non-local conserved currents coming
from the classical sigma model (the limit of $g$ small) are not
conserved once loop corrections are included.  This certainly does not
prove the model is not integrable, because it is possible that some or
all of the classical conserved currents can be modified so that they
are conserved in the full theory.

A simple Lie group has only one factor. Thus the symmetric spaces with
$H$ simple are $O(N)/O(N-1)$, $SU(2N)/Sp(2N)$ and $SU(N)/SO(N)$, and
the principal chiral models $H\times H/H$.  All of these models are
indeed integrable. However, the $O(2P)/O(P)\times O(P)$ models are
also integrable, but $H$ is not simple! Thus the suggestion of
\cite{Abdalla} is in not true here. It is not clear whether this is a
fluke of this model, or other symmetric-space sigma models are
integrable as well. It would be most interesting to construct the
non-local conserved currents here explicitly, to understand how
they remain conserved even in the full theory. Some interesting results
for the classical model were found in \cite{evans2}, but they await
generalization to the quantum case.

The $S$ matrices described above are all what are known as 
rational solutions of the Yang-Baxter equation. This means the
$S$ matrices are rational functions of the rapidity (except for
the prefactor). Yangians are all associated with rational solutions
of the Yang-Baxter equation, so the results described above
certainly imply that there is a Yangian
symmetry in all the integrable sigma models. In fact, this is the
reason for the extra particles in the models with $O(2P)$
symmetry. The representations of the Yangian of $O(2P)$ are larger than
that of its subalgebra $O(2P)$. The particles at a given mass are in a
reducible representation of $O(2P)$, but in an irreducible
representation of the Yangian. This poses an interesting question: is
there any way of telling which representations of the Yangian yield
the particles and $S$ matrices for an integrable field theory? And if
so, what are these theories? Unfortunately, the technology of Yangians
does not seem developed enough yet to answer these questions.

The results of \cite{mecoset} discussed in section 5 do suggest an
alternate approach to finding integrability in sigma models.  It is
much easier to look for conserved currents in perturbed conformal
field theory than it is in sigma models.  For example, it was
noted in \cite{mecoset} that there are (at least to lowest order in
perturbation theory) conserved non-local currents in the
$SU(N)_k/SU(N-1)_k\times U(1)$ coset models perturbed by the operator
${\cal O}_\sigma$. Thus one expect these currents to persist in the
$CP^{N-1}$ sigma model, obtained by taking $k\to\infty$.  Even if
these currents do remain in this limit, this does not prove the
$CP^{N-1}$ models are integrable. However, at the very least it would
indicate that interesting behavior in the sigma models is still lying
yonder.

\bigskip\bigskip 

My work is supported by a DOE OJI Award, a Sloan
Foundation Fellowship, and by NSF grant DMR-9802813.

\appendix

\section{Kernels and identities}

\subsection{$SU(N)$}

One set of kernels I use comes from the prefactors of the
$S$ matrices. These kernels are defined as
\begin{eqnarray*}
A^{(N)}_{ab}(\beta)&=&2\pi\delta_{ab}\delta(\beta)
+i\frac{d}{d\beta} \ln X^{(N)}_{ab}(\beta)\\
Y^{(N)}_{ab} (\beta)&\equiv& {-i}\frac{d}{d\beta} \ln F_{GN}^{ab}(\beta)\\ 
Z^{(N)}_{ab} (\beta)&\equiv& {-i}\frac{d}{d\beta} \ln F^{ab}(\beta) \\
\zeta^{(N,k)}_{ab} (\beta)&\equiv& {-i}\frac{d}{d\beta} \ln F^{ab;k}(\beta)
\end{eqnarray*}
These kernels arise in the prefactors of
the $SU(N)$ parafermion theories, the
$SU(N)$ Gross-Neveu models, the
$SU(N)/SO(N)$ sigma models, and the $SU(N)_s/SO(N)_{2s}$ perturbed
coset models, respectively.  The reason for the extra factor in the 
definition of $A^{(N)}_{ab}$ will become apparent below.
The kernel appearing in vector-vector
scattering is defined as $Y^{(N)}=Y^{(N)}_{11}$.  It is most
useful to give the kernels in Fourier space. To make the equations
look a little nicer, I define the Fourier transform with normalization
\begin{equation}
\widehat{f}(\omega) =
\frac{N}{\pi}
\int_{-\infty}^\infty \frac{d\omega}{2\pi}\ e^{Ni\omega\beta/\pi} f(\beta)
\label{fourier}
\end{equation}
I use this definition of Fourier transformation for any kernel in
a model with $SU(N)$ symmetry.
A fact useful for obtaining the TBA equations in this paper
is that if $\widehat{f}(\omega)=1/\cosh(\omega)$ 
then $f(\beta) = N/(4\pi\cosh(N\beta/2))$.

For the Gross-Neveu models, by using the $S$ matrices in \cite{natan,sunGN}
one finds after some 
after some manipulation \cite{Hollo}
\begin{equation}
\widehat{Y}^{(N)}_{ab}(\omega)= 
\delta_{ab}-
e^{|\omega|}\frac{\sinh((N-a)\omega)\sinh(b\omega)}
{\sinh(N\omega)\sinh(\omega)}
\label{gamm2}
\end{equation}
for $a\ge b$, with $Y^{(N)}_{ab} = Y^{(N)}_{ba} $ To find
the kernels $F^{ab}$ appearing in the $SU(N)/SO(N)$ sigma models
requires even more work. Using the results of \cite{metheta} for the
$S$ matrices, I find
\begin{equation}
\widehat{Z}^{(N)}_{ab}(\omega)= 
\delta_{ab}-
e^{-|\omega|}\frac{4\cosh(\omega)\sinh((N-a)\omega)\sinh(b\omega)}
{\sinh(N\omega)}
\label{eta}
\end{equation}
Notice how the Fourier transforms are related:
$$\widehat{Z}^{(N)}_{ab}(\omega)-\delta_{ab} = 
e^{-2\omega}\sinh(2\omega)\left(\widehat{Y}^{(N)}_{ab}(\omega)-
\delta_{ab}\right).$$
This relation is useful in proving various identities.
Finally, for the perturbed coset models, one has
\begin{eqnarray}
\nonumber
{\zeta}^{(N,s)}_{ab}(\omega)= 
\delta_{ab}-
\frac{4\cosh(\omega)\sinh((N-a)\omega)\sinh(b\omega)\sinh((s-1)\omega)}
{\sinh(N\omega)\sinh(s\omega)}
\label{zeta}
\end{eqnarray}
Note that $Z^{(N)} =\zeta^{(N,\infty)}$, in accord with the idea
in \cite{FZ,mecoset} that the sigma models can be obtained as the limit
of perturbed coset models.

The kernel $A^{(s)}_{ab}$ arises in several places.  The functions
$X^{(N)}_{ab}$ are the $S$ matrix elements 
for the $SU(N)$ parafermion theories, and
appears as part of the prefactor in the Gross-Neveu and $SU(N)/SO(N)$
sigma models.  $A^{(s)}_{ab}$ also arises in the Bethe ansatz
diagonalization. It is
\begin{equation}
\widehat{A}^{(s)}_{jl} (\omega)= 
\frac{2\sinh((s-j)\omega) \cosh(\omega)\sinh(l\omega)}
{\sinh(\omega)\sinh(s\omega)}
\label{Ajl}
\end{equation}
for $j\ge l$,
with $A_{lj}^{(s)}\equiv A^{(s)}_{jl}$. 
Other kernels arising in the Bethe ansatz diagonalization
are 
\begin{equation}
\widehat{\sigma}^{(s)}_j (\omega)= 
\frac{\sinh((s-j)\omega)}{\sinh(s\omega)}
\label{sigmaj}
\end{equation}
in the Gross-Neveu models, and
\begin{equation}
\widehat{\tau}^{(s)}_j (\omega)=
\frac{2\sinh((s-j)\omega)\cosh(\omega)}{\sinh(s\omega)}
 - \delta_{j1}
\label{tauj}
\end{equation}
in the sigma models.
Notice that $\tau$ and $\sigma$ are related via (\ref{tausigma}).
Naively, this seems to imply
$\widehat\tau^{(s)}_j (\omega) = 2\cosh(\omega) 
\widehat\sigma^{(s)}_j(\omega)$, but this is not quite true.
The $\delta_{j1}$ appears in (\ref{tauj}) after a careful analysis
of the Fourier transforms; note that the correct forms vanish as 
$\omega\to\infty$.

The inverses of the matrices $A^{(s)}_{jl}$ are very useful.
By using the Fourier transforms, it is simple to derive
the identity
$$\sum_{k=1}^{s-1} K^{(s)}_{jk}*A^{(s)}_{kl}(\beta) = 
\delta(\beta)\delta_{jl}$$
where
\begin{equation}
\widehat{K}^{(s)}_{jl}(\omega) =
\delta_{jl} - \frac{I^{(s)}_{jl}}{2\cosh(\omega)}
\label{Kjl}
\end{equation}
where $I^{(s)}_{jl}$ is the incidence matrix for the algebra
$SU(s)$, defined in (\ref{inc}).
More generally, the incidence matrix for a
simply-laced Lie algebra is twice the identity minus the Cartan
matrix, and is conveniently pictured by the Dynkin diagram.
I denote the incidence matrix for $SO(2s)$ as ${\cal I}^{(s)}$.
Other useful identities are
$$\sum_{l=1}^{s-1} K^{(s)}_{jl} * \sigma^{(s)}_l (\beta)=  
\delta_{j1}\frac{N}{4\pi\cosh(N\beta/2)}$$
and
$$\sum_{l=1}^{s-1} K^{(s)}_{jl} * \tau^{(s)}_l (\beta)=
\delta_{j2}\frac{N}{4\pi\cosh(N\beta/2)}.$$ 
Useful identities involving the $S$ matrix prefactors are
$$\widehat{Y}^{(N)}_{ab}(\omega)-\delta_{ab}=
\widehat{A}^{(N)}_{ab}(\omega)
\left(\frac{\sigma_1^{(\infty)}(\omega)}{2\cosh(\omega)} - 1\right)$$
and
$$\widehat{Z}^{(N)}_{ab}(\omega)-\delta_{ab}=
\widehat{A}^{(N)}_{ab}(\omega)
\left(\frac{\tau_2^{(\infty)}(\omega)}{2\cosh(\omega)} - 1\right)$$
The extra $\delta_{j1}$
in (\ref{tauj}) is crucial to obtaining the right identities.

\subsection{$O(2P)$}

The $S$ matrices and prefactors for the $O(2P)$ Gross-Neveu models are
given in \cite{KT}, and those for the $O(2P)/O(P)\times O(P)$ in
\cite{metheta}. The kernels are defined as ${\cal Y}^{(P)}_{ab}$ and
${\cal Z}^{(P)}_{ab}$ respectively.  The Fourier transform used below
is that of (\ref{fourier}) with $N$ replaced with $2P-2$.

For $a,b =1\dots P-2$, the Gross-Neveu
kernels are closely related to the $SU(2P-2)$ 
kernels, namely
\begin{eqnarray*}
\nonumber
\widehat{\cal Y}^{(P)}_{ab}(\omega) &=& 
\widehat{Y}^{(2P-2)}_{ab}(\omega) + \widehat{Y}^{(2P-2)}_{2P-2-a\ b}(\omega)\\
&=& 
\delta_{ab}-
e^{|\omega|}\frac{\cosh((P-1-a)\omega)\sinh(b\omega)}
{\cosh((P-1)\omega)\sinh(\omega)}
\label{yon}
\end{eqnarray*}
for $a\ge b$, with 
$\widehat{\cal Y}^{(P)}_{ba} =\widehat{\cal Y}^{(2P-2)}_{ab}$.
For those involving the spinor representations $s$ and $\bar s$ (the nodes
labelled $P-1$ and $P$), the kernels are
\begin{eqnarray*}
\widehat{\cal Y}^{(P)}_{PP} &=&\widehat{\cal Y}^{(P)}_{P-1\,P-1} = 
\widehat{\cal Y}^{(P)}_{P\,P-1} =
1-e^{|\omega|}\frac{\sinh(P\omega)}{2\sinh(2\omega)\cosh((P-1)\omega)}\\
\widehat{\cal Y}^{(P)}_{aP} &=&\widehat{\cal Y}^{(P)}_{a\,P-1} = 
-e^{|\omega|}\frac{\sinh(a\omega)}{2\sinh(\omega)\cosh((P-1)\omega)}
\end{eqnarray*}
where in the latter $a=1\dots P-2$.

The result of \cite{BRii,BRiii} for the Bethe equations
for $O(2P-2)$
says that the equation for the eigenvalue (\ref{evGN}) and 
the first Bethe equation (\ref{betGN}) are 
\begin{equation}
2\pi P_{a,0} (\beta) = {m_a}\cosh\beta + 
\sum_{b=1}^{P}Y^{(P)}_{ab}*\rho_{b,0}(\beta) - 
\sum_{l=1}^\infty
\sigma^{(\infty)}_{l} * \widetilde{\rho}_{a,l}(\beta).
\label{betGNON}
\end{equation}
These are virtually identical to those for $SU(2P-2)$, with
$Y_{ab}^{(N)}$ replaced by $Y_{ab}^{(2P-2)}$. In particular,
the kernel $\sigma^{(2P-2)}$ is still given by (\ref{sigmaj}).
The other Bethe equations are now
\begin{equation}
2\pi {\rho}_{a,j}(\beta) = 
\sigma_j^{(\infty)} * \rho_{a,0}(\beta) - \sum_{b=1}^{P}\sum_{l=1}^\infty 
A_{jl}^{(\infty)} * {\cal K}_{ab}^{(P)} * 
\widetilde\rho_{b,l}(\beta)
\label{betGNk}
\end{equation}
where
\begin{equation}
\widehat{\cal K}^{(s)}_{jl}(\omega) =
\left( \delta_{jl} - \frac{{\cal I}^{(s)}_{P-j\ P-l}}
{2\cosh(\omega)}\right)
\label{Kjk}
\end{equation}
where $I^{(P)}_{jl}$ is the incidence matrix for the algebra
$SO(2P)$, defined above in (\ref{incSO}). The reason for the $P-j$ and $P-l$
indices
is that above it was convenient above 
to define the spinor nodes as $0$ and $1$,
whereas here I have defined them as $P$ and $P-1$.

The proof of the TBA equations is now basically identical to that
done for the $SU(N)$ Gross-Neveu model. The reason is that the kernels
here satisfy basically the same identities as the $SU(N)$ case. Namely,
one can define the matrix inverse ${\cal A}$ of ${\cal K}$, just like
$A$ is the inverse of $K$.
One finds that
$$-i\frac{d}{d\beta} \ln {\cal X}^{ab}(\beta)
=\delta_{ab}\delta(\beta)-{\cal A}^{(P)}_{ab}(\beta)$$
Then
$$\widehat{\cal Y}^{(P)}_{ab}(\omega)-\delta_{ab}=
\widehat{\cal A}^{(P)}_{ab}(\omega)
\left(\frac{\widehat{\sigma}_1^{(\infty)}(\omega)}{2\cosh(\omega)} - 1\right)$$
Using this and the identities in the first appendix gives the $O(2P)$
Gross-Neveu TBA equations in (\ref{tbaGNON}).

For the $O(2P)/O(P)\times O(P)$ models, the proof is the same as
for the $SU(N)/SO(N)$ models. The only new identity needed is
$$\widehat{\cal Z}^{(P)}_{ab}(\omega)-\delta_{ab}= \widehat{\cal
A}^{(P)}_{ab}(\omega)
\left(\frac{\widehat{\tau}_2^{(\infty)}(\omega)}{2\cosh(\omega)} 
- 1\right)$$ From the
prefactor given in \cite{metheta}, it follows that this identity
holds for $a=b=1$. However, I have not been able to prove it in
general. The reason is that the $S$ matrices for particles in the
representations $2\mu_s$ and $2\mu_{\bar s}$ are not known explicitly,
so it has not been possible to work out the prefactors involving these
particles. However, I have checked that if they obey the above
identity, then they are consistent with the massive TBA.
By consistent, I mean that the TBA equations are the same as in
the massive case with only different asymptotic conditions, so that
the perturbative expansion of the free energy 
is the same for $\theta=0$ and
$\theta=\pi$. I have also checked this consistency for the energy
at zero temperature in a magnetic field, extending the analysis of
\cite{metheta} to the particles in representations $\mu_s$ for the
massless case and $2\mu_s$ in the massive case.

As a tangential note, the $O(2P)/O(2P-1)$ sigma models are integrable as
well \cite{ZZ}. Their spectrum consists of a single multiplet of
$2P$ particles in the vector representation, with no bound states.  
The TBA equations are very similar \cite{mecoset}, but
$a$ in $\rho_{a0}$ can only be $1$. The other $\epsilon_{aj}$ still
have $a=1\dots P$.
Because there are no bound states, the prefactor ${\cal F}^{11}$ is
not the same is in the Gross-Neveu models:
${\cal X}(\beta)$ needs to be removed from the prefactor \cite{ZZ}.
The kernel appearing in the TBA equations is therefore
${\cal Y}^{(2P)}_{11}(\beta) 
-\delta_{ab}\delta(\beta)+ {\cal A}^{(2P)}_{11}(\beta)$.  
Using this with the
above Bethe equations gives the TBA equations given in \cite{mecoset}.

\renewcommand{\baselinestretch}{1}

\end{document}